\documentclass[
reprint,
superscriptaddress,
amsmath,
amssymb,
aps,
longbibliography,
pra,
showpacs,
floatfix
]{revtex4-1}

\usepackage[mathlines]{lineno}
\usepackage{graphicx}
\usepackage[margin=1in]{geometry} 
\usepackage{graphicx}
\usepackage[table]{xcolor}

\usepackage{tabularx}
\usepackage{multirow}
\usepackage{dcolumn} 
\newcolumntype{d}[1]{D{.}{.}{#1}}
\newcolumntype{L}[1]{>{\raggedright\arraybackslash}p{#1}}
\newcolumntype{C}[1]{>{\centering\arraybackslash}p{#1}}
\newcolumntype{R}[1]{>{\raggedleft\arraybackslash}p{#1}}
\usepackage{booktabs}

\usepackage{comment}
\usepackage{enumitem}

\usepackage[hidelinks]{hyperref}
\definecolor{color1}{rgb}{0,0.25,0.70}
\hypersetup{colorlinks=true, 
    linkcolor={color1},
    citecolor={color1}, 
    urlcolor={color1}
}

\usepackage[]{xcolor}
\usepackage{soul}
\usepackage{cancel,siunitx}

\renewcommand{\vec}{\bm}
\renewcommand{\epsilon}{\varepsilon}

\usepackage{physics}
\usepackage{bm}

\graphicspath{ {./figures/} }

\begin{document}

\preprint{APS/123-QED}

\title{Molecular van der Waals fluids in cavity quantum electrodynamics}

\author{John P. Philbin$^{\P}$}
 \email{jphilbin01@gmail.com}
 \affiliation{Harvard John A. Paulson School of Engineering and Applied Sciences, Harvard University, Cambridge, MA 02138, USA}
 \affiliation{College of Letters and Science, University of California, Los Angeles, CA 90095, USA%
}

\author{Tor S. Haugland$^{\P}$}
 \affiliation{Department of Chemistry, Norwegian University of Science and Technology, 7491 Trondheim, Norway%
}

\author{Tushar K. Ghosh$^{\P}$}
 \affiliation{Department of Chemistry, Purdue University, West Lafayette, IN 47907, USA%
}

\author{Enrico Ronca}
 \affiliation{Dipartimento di Chimica, Biologia e Biotecnologie, Università degli Studi di Perugia, Via Elce di Sotto, 8, 06123, Perugia, Italy}
 \affiliation{Max Planck Institute for the Structure and Dynamics of Matter and Center Free-Electron Laser Science, Luruper Chaussee 149, 22761 Hamburg, Germany%
}

\author{Ming Chen}
 \email{chen4116@purdue.edu}
 \affiliation{Department of Chemistry, Purdue University, West Lafayette, IN 47907, USA%
}

\author{Prineha Narang}
 \email{prineha@ucla.edu}
 \affiliation{Harvard John A. Paulson School of Engineering and Applied Sciences, Harvard University, Cambridge, MA 02138, USA}
 \affiliation{College of Letters and Science, University of California, Los Angeles, CA 90095, USA
}

\author{Henrik Koch}
 \email{henrik.koch@sns.it}
 \affiliation{Department of Chemistry, Norwegian University of Science and Technology, 7491 Trondheim, Norway}
 \affiliation{Scuola Normale Superiore, Piazza dei Cavalieri, 7, 56124 Pisa, Italy%
}

\thanks{Denotes equal contribution}

\date{\today}

\begin{abstract}

Intermolecular van der Waals interactions are central to chemical and physical phenomena ranging from biomolecule binding to soft-matter phase transitions. However, there are currently very limited approaches to manipulate van der Waals interactions. In this work, we demonstrate that strong light-matter coupling can be used to tune van der Waals interactions, and, thus, control the thermodynamic properties of many-molecule systems. Our analyses reveal orientation dependent single molecule energies and interaction energies for van der Waals molecules (for example, H$_{2}$). For example, we find intermolecular interactions that depend on the distance between the molecules $R$ as $R^{-3}$ and $R^{0}$. Moreover, we employ non-perturbative \textit{ab initio} cavity quantum electrodynamics calculations to develop machine learning-based interaction potentials for molecules inside optical cavities. By simulating systems ranging from $12$ H$_2$ to $144$ H$_2$ molecules, we demonstrate that strong light-matter coupling can tune the structural and thermodynamic properties of molecular fluids. In particular, we observe varying degrees of orientational order as a consequence of cavity-modified interactions, and we explain how quantum nuclear effects, light-matter coupling strengths, number of cavity modes, molecular anisotropies, and system size all impact the extent of orientational order. These simulations and analyses demonstrate both local and collective effects induced by strong light-matter coupling and open new paths for controlling the properties of molecular clusters.

\end{abstract}

\maketitle

Van der Waals interactions are ubiquitous in chemistry and physics, playing important roles in diverse scientific fields ranging from DNA base stacking to 2D material interlayer interactions.\cite{Hobza2002,Novoselov2016,Sternbach2021} There has been a long history of attempting to elucidate the origin of van der Waals interactions;\cite{Maitland1981,Stone2013} the first quantum mechanical derivation was performed by London in the 1930s using second-order perturbation theory.\cite{London1937} London found that two molecules that do not have permanent dipoles (e.g. H$_2$), which we refer to as van der Waals molecules, have an attractive interaction between them that scales with the distance between the molecules $R$ as $R^{-6}$.\cite{London1937} This $R^{-6}$ attractive force is commonly used as the long-distance asymptotic form of van der Waals interactions in many force fields and to correct van der Waals interactions in \textit{ab initio} calculations, which have both achieved great successes in modeling thermodynamic properties in a variety of systems.\cite{Halgren1992,Grimme2010} Despite van der Waals interactions being central to many properties of molecular and condensed matter systems, limited approaches have been proposed to manipulate intermolecular van der Waals interactions. However, applied electromagnetic fields have been shown to modify van der Waals interactions between atoms and molecules,\cite{Thirunamachandran1980,Milonni1996,Sherkunov2009,Fiscelli2020} and Haugland et al.\cite{Haugland2021} recently showed numerically that van der Waals interactions are significantly altered by strong light-matter coupling in optical cavities. These studies open the possibility of controlling the properties and structure of molecular fluids by tuning the light-matter coupling parameters, the coupling strength and frequency.

\begin{figure*}[htbp]
    \centering{}
    \includegraphics[width=16.0cm]{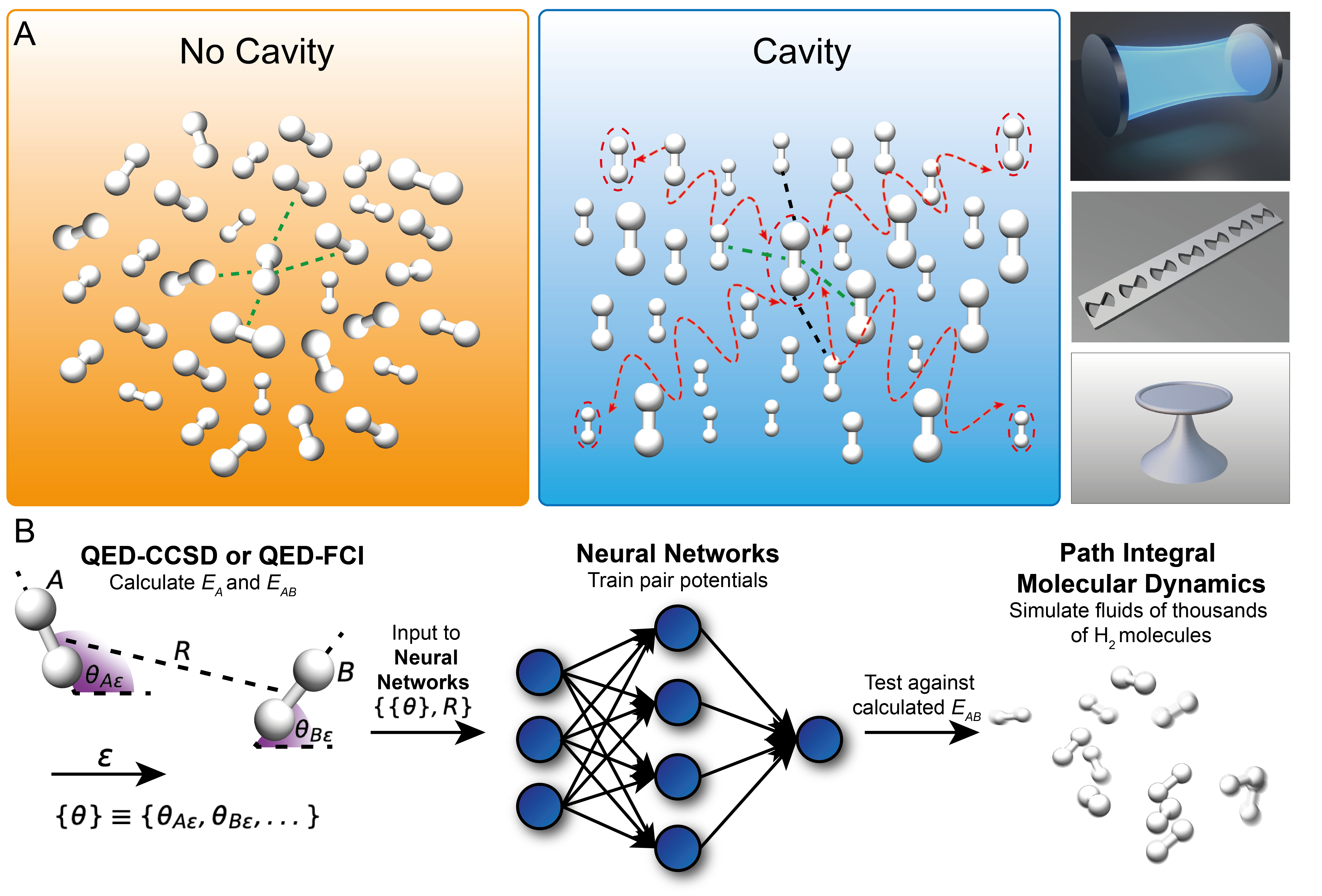}
    \caption{\label{fig:fig-1-cartoon}(A) Schematic representation of the findings from our simulations of a fluid of H$_2$ molecules outside and inside a cavity. Specifically, orientational order can be observed inside a cavity whereas the H$_2$ molecules can rotate freely outside of a cavity. The dashed lines represent the different intermolecular interaction length scales outside and inside a cavity. (B) Diagram describing the computational workflow used in this work. \textit{Ab initio} cavity QED energies and corresponding symmetry preserving features (see Fig.~S3, Table~S3 and Section~SIV.A.1 for details of symmetry preserving features) of many $2$H$_2$ configurations are used to develop neural network-based intermolecular pair potentials capable of being utilized in path integral molecular dynamics simulations of fluids of H$_2$ molecules.}
\end{figure*}

The goal of this work is to understand how the structure of molecular van der Waals fluids can be modulated using enhanced vacuum electromagnetic quantum fields, and we focus on the impact that a single strongly coupled photon mode can have on the properties of a model molecular van der Waals fluids. To this end, we leverage recent developments in cavity quantum electrodynamics (QED) simulations and neural network pair potentials to simulate molecular fluids of H$_2$ molecules strongly coupled to a single photon mode (Fig.~\ref{fig:fig-1-cartoon}). By analyzing how cavity-modified single molecule energies and cavity-mediated intermolecular interactions depend on the orientation of the H$_2$ molecules both relative to the cavity polarization vector and relative to one another, we can explain how cavities impact the structure and orientational order of molecular van der Waals fluids. The findings reported herein should readily be transferable to other molecules and light-matter regimes (e.g. vibrational polaritons) given the generality of the cavity QED Hamiltonian used in this work.\cite{RibeiroChemSci2018,Rivera2019,ThomasScience2019,Li2020a,Garcia-Vidal2021,Li2021} We also discuss how the light-matter coupling strength, number of cavity modes, temperature, anisotropic polarizabilities of molecules, quantum nuclear effects, and molecular concentrations can all impact the extent of orientational order observed in any particular cavity QED experiment.\cite{Vahala2003,deLiberato2017,Joseph2021,Fukushima2022,Sandeep2022}  

In molecular dynamics (MD) simulations, the nuclei move along electronic potential energy surfaces. In the cavity case, where the photon contributions are added, these surfaces have been termed polaritonic potential energy surfaces.\cite{GalegoPhysRevX2015,Lacombe2019,Fregoni2022} In both cases, the total potential energy of $N$ H$_2$ molecules can be calculated as a many-body expansion,
\begin{equation}
    E_{\text{total}} = \sum_{A}E_{A} + \sum_{\left\langle A,B \right\rangle}E_{AB} + {\sum_{\left\langle A,B,C \right\rangle}E_{ABC}} + ...,
    \label{eq:total-energy}
\end{equation}
where $E_A$ represents the single-molecule energies, $E_{AB}$ represents the intermolecular interaction energies between all unique pairs of molecules, and so on for higher-body terms. In this work, we focus on contributions to the total energy in Eq.~\ref{eq:total-energy} arising from at most two-body interactions. The three-body and higher-body terms are significantly smaller than the two-body interactions per interaction, see the Supplementary Information (SI) for details. Outside the cavity, the one-body term does not depend on the orientation of the H$_2$ molecule. On the other hand, inside the cavity, the molecule-field interaction causes the one-body energies to depend on the orientation of the H$_2$ molecules with respect to the optical cavity polarization vector, $\vec{\epsilon}$. Furthermore, the two-body energies depends on the orientation between the two molecules as well as their orientation relative to the field as a consequence of the anisotropic polarizability of H$_2$ molecules, in contrast to isotropic polarizabilities of atoms.\cite{Thirunamachandran1980,Milonni1996,Sherkunov2009,Fiscelli2020} 

\begin{figure*}[htbp]
    \centering{}\includegraphics[width=\textwidth]{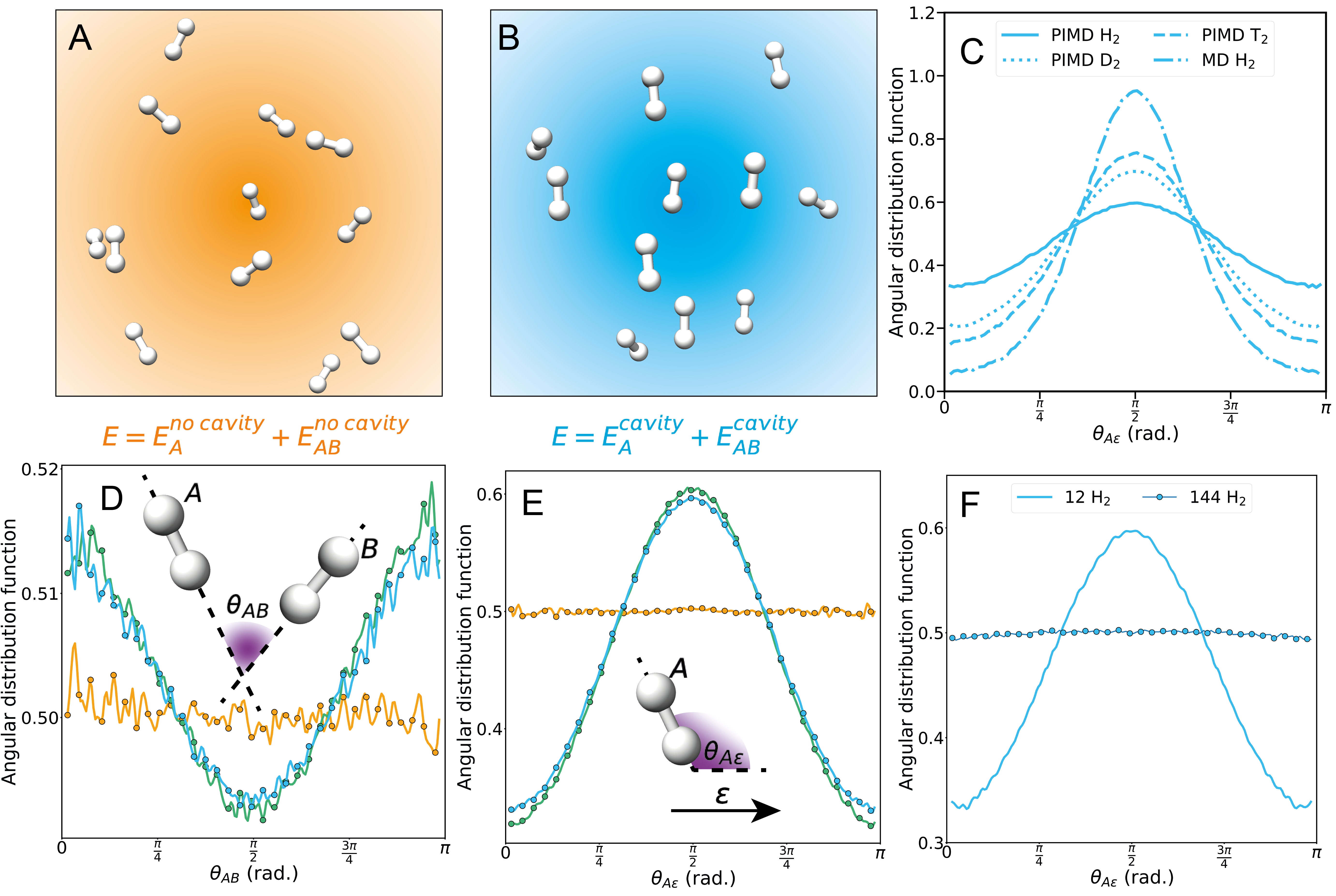} \caption{\label{fig:fig-2-MD}(A-B) Snapshots taken at thermal equilibrium from molecular dynamic (MD) simulations in the case of (A) no cavity (orange) and (B) cavity-modified one-body and two-body terms (blue). (C) The impact of quantum nuclear effects are demonstrated by comparing the molecular bond axis to cavity polarization vector ($\theta_{A\epsilon}$), angular probability distribution function, $P\left(\theta_{A\epsilon}\right)$ for path integral molecular dynamics (PIMD) simulations of H$_2$, D$_2$, T$_2$, and a classical MD simulation of H$_2$. (D) Molecular bond axis of molecule $A$ to molecular bond axis of molecule $B$ ($\theta_{AB}$) angular probability distribution function, $P\left(\theta_{AB}\right)$ and (E) $P\left(\theta_{A\epsilon}\right)$ are shown for PIMD simulations for no cavity (orange), cavity (blue), and cavity-modified one-body term but no cavity two-body term (green) cases. (F) $P\left(\theta_{A\epsilon}\right)$ is shown for two different PIMD simulations containing different numbers of H$_2$ molecules within the same cavity volume (i.e. changing the molecular density). All PIMD simulations shown in this figure were performed using neural networks trained with CCSD (no cavity) or QED-CCSD-12-SD1 with $\lambda=0.1$ a.u. (cavity) calculated energies. All entropic contributions to angle distribution functions are removed.}
\end{figure*}

We calculate $E_A$ and $E_{AB}$ by solving the Schrödinger equation for the cavity QED Hamiltonian in the dipole approximation with a single photon mode using accurate coupled cluster (QED-CCSD-12-SD1) and near exact full configuration interaction (QED-FCI-5).\cite{Haugland2020} Our single photon mode has a coupling constant of $\lambda=0.1$~a.u. and energy of $\hbar\omega_c=13.6$~eV unless specified otherwise. This coupling constant is rather large as it corresponds to the coupling of at least $5$ independent modes where each has an effective volume of $0.9$~nm$^3$. We detail below how the cavity-modified local interactions and cavity-induced collective effects depend on $\lambda$.
More than $100,000$ H$_2$ dimer configurations are used as inputs to a fully-connected neural network that serves as our intermolecular pair potential, which is trained and tested against the calculated energies. The trained potential energy functions were carefully tested, and, in the SI, we demonstrate that our machine learning models are fully capable of reproducing the potential energy surfaces. In Fig.~\ref{fig:fig-1-cartoon}B, we show the computational workflow used in this work schematically. In this study, we focus on path integral molecular dynamics (PIMD) simulations in order to account for quantum nuclear effects. Our PIMD simulations of fluids of H$_2$ molecules were performed with a fixed number of molecules ($N$), temperature ($T$), and volume ($V$). All PIMD simulations presented herein were performed with a molecular density of $13$~molecules per nm$^3$, temperature of $70$~K, and $N=12$ unless otherwise specified. More details on the simulations, including comparisons of QED-CCSD-12-SD1 with QED-FCI-5, comparisons of MD with PIMD, and additional parameter regimes (e.g. smaller $\lambda$ values), are provided in the SI.

\begin{figure*}[htbp]
    \centering{}
    \includegraphics[width=\textwidth]{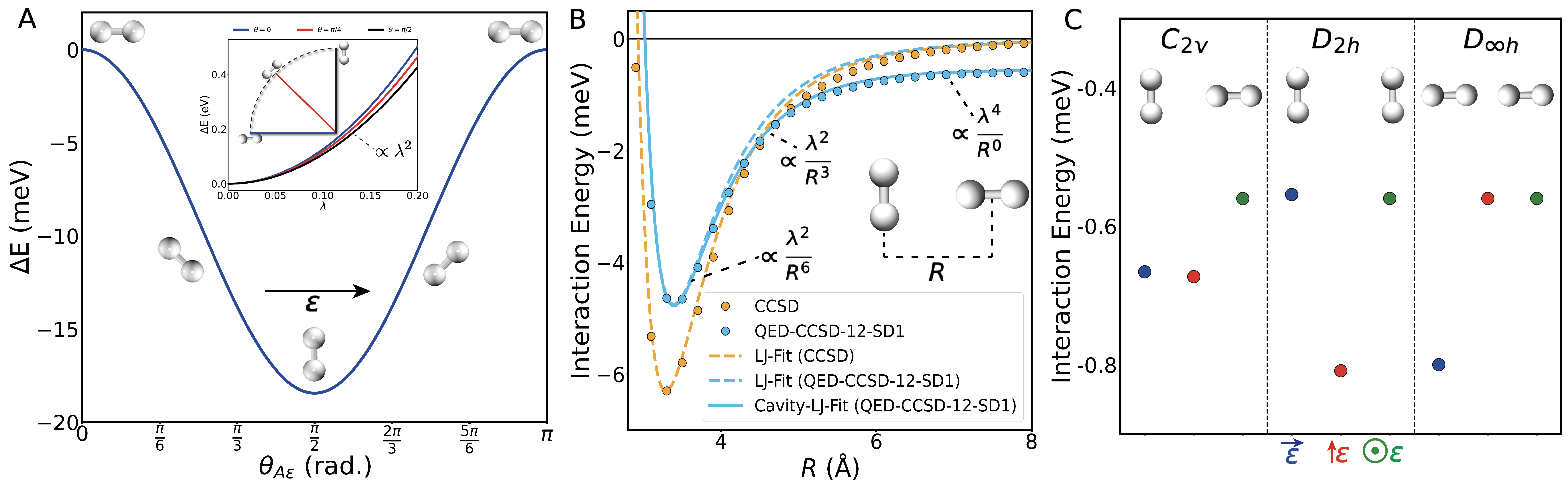}
    \caption{\label{fig:fig-3-PT}(A) Energy difference, $\Delta E$, between a single H$_2$ molecule inside a cavity aligned perfectly along the cavity polarization vector, $\vec{\epsilon}$, and different angles relative to the cavity polarization vector. The inset shows the energy of a single molecule within a cavity increases with $\lambda^2$. (B) Intermolecular interaction energies, $E_{AB}$, and fits to a Lennard-Jones type potential given by Eq.~\ref{eq:E_AB_no_cavity} (dashed lines) and cavity-modified Lennard Jones type potential given by Eq.~\ref{eq:E_AB_cavity} (solid line). (C) Intermolecular interaction energies, $E_{AB}$, at $25$~{\AA} for various high symmetry molecular orientations and cavity polarizations. All calculations shown in this figure were performed using QED-CCSD-12-SD1 with $\lambda=0.1$~a.u.}
\end{figure*}

The structural properties of the molecular van der Waals fluids are analyzed using PIMD simulation trajectories. In Fig.~\ref{fig:fig-2-MD}, we summarize the main findings of our PIMD and classical MD simulations. Fig.~\ref{fig:fig-2-MD}A and Fig.~\ref{fig:fig-2-MD}B show representative thermal equilibrium configurations for the no cavity (orange) and cavity (blue) scenarios, respectively. The impact of the cavity-modified interactions are observable in the orientational order of the H$_2$ molecules both relative to the cavity polarization vector ($\theta_{A\epsilon}$, Figs.~\ref{fig:fig-2-MD}C, E and F) and relative to other H$_2$ molecules ($\theta_{AB}$, Fig.~\ref{fig:fig-2-MD}D). Specifically, Figs.~\ref{fig:fig-2-MD}C-F show that the cavity-modified energies enhance the probability of finding two molecules oriented parallel to one another (i.e. $\theta_{AB}=0,\pi$) and perpendicular to the cavity polarization vector (i.e. $\theta_{A\epsilon}=\frac{\pi}{2}$). However, the extent of this orientational order depends on many factors, including the magnitude of quantum nuclear effects, the light-matter coupling strengths, molecular anisotropies, and number of molecules. To elucidate the importance of quantum nuclear effects, we compare the orientational order observed in PIMD simulations of H$_2$, D$_2$, and T$_2$ with a classical MD simulation of H$_2$ in Fig.~\ref{fig:fig-2-MD}C; the degree of orientational order monotonically increases upon increasing the molecular masses from H$_2$ to D$_2$ to T$_2$ (which reduces quantum nuclear effects) and is further enhanced when quantum nuclear effects are completely removed as in the classical MD simulation. Next, in Figs.~\ref{fig:fig-2-MD}D-F, we show how cavity-modified one-body energies and two-body intermolecular energies each impact the orientational order. Fig.~\ref{fig:fig-2-MD}D and Fig.~\ref{fig:fig-2-MD}E demonstrate that the cavity-modified one-body energies are the dominant driver of the orientational order for the case of $12$ H$_2$ molecules. The orange lines in Figs.~\ref{fig:fig-2-MD}D,E show that the H$_2$ molecules have no preferred orientation axis outside the cavity, consistent with the global rotational symmetry of the electronic and nuclear Hamiltonian in absence of the cavity. However, the presence of the bilinear coupling and dipole self-energy terms break this symmetry such that H$_2$ molecules prefer to orient their bond axis in specific orientations relative to the cavity polarization vector and relative to one another. In particular, the dipole self-energy term outcompetes the bilinear coupling term and is responsible for the $12$ molecule simulations preferentially aligning perpendicular to the cavity polarization vector (Fig.~\ref{fig:fig-3-PT}A). However, Figs.~\ref{fig:fig-2-MD}E,F demonstrate that the cavity-modified one-body energies lead to this perpendicular alignment whereas the cavity-modified two-body intermolecular interactions attempt to align the molecules parallel to the cavity polarization vector. Specifically, the green line in Fig.~\ref{fig:fig-2-MD}E shows that the cavity-modified one-body term causes H$_2$ molecules to preferentially align perpendicular to the cavity polarization vector (i.e. $\theta_{A\epsilon}=\frac{\pi}{2}$), and the inclusion of cavity-modified two-body interactions begins to counteract this effect as seen in the blue line in Fig.~\ref{fig:fig-2-MD}E reducing the orientational alignment. This effect of the two-body interactions causing the H$_2$ molecules to preferentially align parallel to the cavity polarization vector (i.e. $\theta_{A\epsilon}=0,\pi$) and the collective nature of the cavity-modified intermolecular interactions are highlighted in Fig.~\ref{fig:fig-2-MD}F and Fig.~S13. We find that for a small number of molecules (e.g. $N=12$) the one-body term dominates and the molecules preferentially align perpendicular to the cavity polarization vector, but as $N$ increases to $144$ H$_2$ molecules with a fixed coupling and cavity volume the orientational order is lost due the cavity-modified one-body and two-body effects perfectly canceling one another. Additionally, the extent of orientational order induced by the cavity decreases as the light-matter coupling strength decreases as shown in Fig.~S8 and explained analytically below.

Although we performed non-perturbative \textit{ab initio} cavity QED calculations, perturbation theory can be used to further analyze and explain the major findings of our PIMD and MD simulations. We summarize our key findings here and in Fig.~\ref{fig:fig-3-PT}, and the complete analysis is provided in the SI. The cavity modifications to the one-body energies, $E_A$, results in the H$_2$ molecules aligning their bonds orthogonal to the cavity polarization. This occurs because H$_2$ is most polarizable along its bond axis, and, from perturbation theory, we can obtain an expression for the cavity-modified one-body energy as
\begin{equation}
    E_{A}^{\text{cavity}} \approx E_{A}^{\text{no cavity}} + c\, (\alpha_{\parallel}\cos^2{\theta_{A\epsilon}}+\alpha_{\perp}\sin^2{\theta_{A\epsilon}}),
    \label{eq:1-body-energy}
\end{equation}
where $\alpha_{\parallel}$ and $\alpha_{\perp}$ are the polarizabilities of molecular hydrogen along its bond axis and perpendicular axes, respectively, and $c$ is a positive scalar constant proportional to the molecule-cavity coupling squared (i.e. $c \propto \lambda^2$). Eq.~\ref{eq:1-body-energy} is in agreement with the \textit{ab initio} calculations shown in Fig.~\ref{fig:fig-3-PT}A. Interestingly, the dipole self-energy term increases the energy of a single molecule in a cavity more than the bilinear coupling term decreases the energy (Eq.~S12); thus, the lowest energy orientation of a single molecule in a cavity is such that its most polarizable axis is perpendicular to the cavity polarization vector (or vectors in terms of multimode cavities). 

In terms of the cavity modifications to the two-body energies, Fig.~\ref{fig:fig-3-PT}B shows the intermolecular interaction between two H$_2$ molecules as a function of the center-to-center distance ($R$). The impact of the cavity on this dissociation curve at first glance appears modest, even for the rather large light-matter coupling of $\lambda=0.1$~a.u., but these modifications can impact the structural and thermodynamic properties of molecular van der Waals systems for a few reasons. First, a standard intermolecular van der Waals interaction potential given by
\begin{equation}
    E_{AB}^{\text{no cavity}} = \frac{c_6}{R^{6}} + E_{\text{short-range}},
    \label{eq:E_AB_no_cavity}
\end{equation}
where $E_{\text{short-range}}$ accounts for the short-range repulsion between van der Waals molecules and the $R^{-6}$ term is the usual attractive London dispersion interaction, is not applicable inside an optical cavity (Fig.~\ref{fig:fig-3-PT}B).\cite{Thirunamachandran1980,Milonni1996,Sherkunov2009,Fiscelli2020} A modified interaction potential that includes angle-dependent terms that scale as $R^{-3}$ and $R^0$ is necessary inside an optical cavity such that the interaction between two van der Waals molecules is given by
\begin{equation}
    E_{AB}^{\text{cavity}} = \frac{c_0}{R^{0}} + \frac{c_{3}}{R^{3}} + \frac{c_6}{R^{6}} + E_{\text{short-range}}.
    \label{eq:E_AB_cavity}
\end{equation}
These interactions arise as early as second-order perturbation theory (see SI Eq.~S9).\cite{Thirunamachandran1980} The $R^0$ interaction between a single pair of molecules is rather weak ($c_0 \propto \lambda^4$) as shown in Fig.~\ref{fig:fig-3-PT}C. However, due to its long-range nature, a single molecule interacts with all other molecules, and, thus, the collective effect of this interaction can become large in many-molecule simulations. Importantly, this interaction strength depends on the orientations of both molecular bonds relative to the cavity polarization (Fig.~\ref{fig:fig-3-PT}C). Specifically, the interaction energy is minimized when the molecular bonds of both molecules are parallel to the cavity polarization vector, because the interaction strength of this term is approximately related to the product of the polarizability of each molecule along $\vec{\epsilon}$ ($c_0 \propto \alpha_{A\epsilon} \alpha_{B\epsilon}$). And because $c_0$ is always negative, this $R^0$ intermolecular interaction increases the probability of finding H$_2$ molecules parallel to the cavity polarization vector and decreases the probability to find the molecules perpendicular to the polarization vector (Fig.~\ref{fig:fig-2-MD}E,F). The collective nature of this interaction is demonstrated in Fig.~\ref{fig:fig-2-MD}F and Fig.~S13 where the orientational order depends on the number of H$_2$ molecules for simulations with the same simulation volume but different molecular densities. At $N=144$, the orientational order due to the two-body interactions have become so large that they entirely cancel out the orientational effects from the cavity modified one-body energies that are dominated by dipole self-energy effects for $N=12$ molecules. As $N$ increases further, we expect that the system will completely flip, and instead align parallel to the polarization vector. This is demonstrated in the SI (Fig.~S13), but the number of molecules required ($N\geq1000$) is too large to justify in a realistic system with the coupling we are using currently. Both the cavity-modified $R^{-6}$ and cavity-induced $R^{-3}$ interactions scale with $\lambda^2$ at lowest order. Importantly, the $R^{-3}$ interaction is not a result of the cavity inducing a dipole moment in the H$_2$ molecules but rather an interaction taking place via the cavity mode. As discussed in the SI in more detail, the intermolecular angle and molecule-cavity angle dependencies of the perturbation potential combine to create the orientational order shown throughout Fig.~\ref{fig:fig-2-MD}.

In summary, we have demonstrated that strong light-matter coupling to a single photon mode can have profound impacts on the properties of molecular van der Waals fluids by combining \textit{ab initio} cavity QED calculations with path integral molecular dynamics simulations of many H$_2$ molecules. We found that cavity-modified single molecule and intermolecular interaction energies result in significantly changed molecular orientational order, even in the fluid phase. We look forward to seeing future experimental and theoretical studies that aim to elucidate how processes such as ion and molecular diffusion, intermolecular energy transfer,\cite{Zhong2016,DuChemSci2018,Xiang2020} and chemical reactivity\cite{herrera2016,ThomasScience2019,Yang2021,Li2021a,Simpkins2021,Philbin2022} are impacted by the unique properties of molecular fluids in cavity QED reported here.

\begin{acknowledgments}
We thank Jonathan Curtis, Davis Welakuh, Wenjie Dou, and Rosario R. Riso for helpful discussions. This work was primarily supported by the Department of Energy, Photonics at Thermodynamic Limits Energy Frontier Research Center, under Grant No. DE-SC0019140 and European Research Council under the European Union’s Horizon 2020 Research and Innovation Programme grant agreement No. 101020016. An award of computer time was provided by the INCITE program. This research also used resources of the Oak Ridge Leadership Computing Facility, which is a DOE Office of Science User Facility supported under Contract DE-AC05-00OR22725 J.P.P. also acknowledges support from the Harvard University Center for the Environment. T.K.G. and M.C. acknowledge support from Purdue startup funding. T.S.H. and H.K. also acknowledges funding from the Research Council of Norway through FRINATEK project 275506. P.N. acknowledges support as a Moore Inventor Fellow through Grant No. GBMF8048 and gratefully acknowledges support from the Gordon and Betty Moore Foundation as well as support from a NSF CAREER Award under Grant No. NSF-ECCS-1944085. E.R acknowledges funding from the European Research Council (ERC) under the European Union’s Horizon Europe Research and Innovation Programme (Grant n. ERC-StG-2021-101040197 - QED-SPIN).
\end{acknowledgments}

\bibliography{cavity} 


%
\clearpage

\renewcommand{\thetable}{Table\hspace{3pt}S\arabic{table}}%
\renewcommand{\thefigure}{Fig.\hspace{3pt}S\arabic{figure}}%
\renewcommand{\theequation}{S\arabic{equation}}%

\setcounter{figure}{0}
\setcounter{equation}{0}
\setcounter{table}{0}

\preprint{APS/123-QED}

\begin{center} \Large\bfseries
Supplementary Information: \\ Molecular van der Waals fluids in cavity quantum electrodynamics
\end{center}

\tableofcontents

\section{\textit{Ab Initio} Calculations}

The Hamiltonian used in the \textit{ab initio} calculations is the single mode Pauli-Fierz Hamiltonian in the length gauge
\begin{align}
    H &= H_e + \lambda \sqrt{\frac{\omega_c}{2}} ((\vec d - \expval{\vec d})\cdot \vec \epsilon)(b + b^\dagger) \label{eq:H_total} \\ \nonumber
    & + \frac{\lambda^2}{2} ((\vec d - \expval{\vec d}) \cdot \vec \epsilon)^2 + \omega_c b^\dagger b,
\end{align}
where $H_e$ is the electronic Hamiltonian, $\lambda$ is the bilinear coupling, $\omega_c$ is the cavity frequency, $\vec{d}$ is the molecular dipole, $\vec{\epsilon}$ is the cavity polarization vector, and $b$ and $b^\dagger$ are the photon annihilation and creation operators, respectively.

All electronic structure calculations are run using an aug-cc-pVDZ basis set. The optical cavity is described by a single linearly polarized mode coupling parameter $\lambda$ is set to $0.1$~a.u. and the cavity energy $\hbar\omega_c$ is $13.6$~eV, unless otherwise specified.

The large value for the coupling is partially justified by the single mode approximation. For cavity-induced changes in the ground state, each cavity mode will to second order in perturbation theory (see Eq.~\ref{eq:1-body-energy-lambda-squared}) enter the energy independently. 
For larger frequencies, the bilinear contribution from each mode cancels part of the dipole self-energy. 
For smaller frequencies compared to electronic excitation energies, we find that only contributions from the dipole self-energy are significant. Therefore, in the low-frequency regime, the coupling from $N_{\rm modes}$ modes is given by an effective coupling $\lambda^2_{\rm eff} \approx N_{\rm modes}\lambda^2$. 

As shown and discussed in Ref.~\cite{Haugland2021}, cavity quantum electrodynamics Hartree-Fock (QED-HF) and current QED density functional theory (QEDFT) implementations do not describe intermolecular forces properly, especially van der Waals interactions in which they fail to predict an attractive interaction between van der Waals molecules. Therefore, we performed the \textit{ab initio} simulations with QED coupled cluster (QED-CCSD-12-SD1) and QED full configuration interaction (QED-FCI).\cite{White2020} QED-CCSD-12-SD1 is an extension of QED-CCSD-1, as described in Ref.~\cite{Haugland2020}, with two-photon excitations. The QED-CCSD-12-SD1 cluster operator is
\begin{equation}
    T = T_1 + T_2 + S_1b^\dagger + S_2 b^\dagger + \gamma_1 b^\dagger + \gamma_2 (b^\dagger)^2,
\end{equation}
where $T_1$ and $T_2$ are singles and doubles electron excitations, $S_1 b^\dagger$ and $S_2 b^\dagger$ are singles and doubles coupled electron-photon excitations, and $\gamma_1 b^\dagger$ and $\gamma_2 (b^\dagger)^2$ are singles and doubles photon excitations. The reference state is QED-HF as described in Ref.~\cite{Haugland2020}. QED-FCI calculations are run with up to five photons (QED-FCI-5) to ensure that the energy with respect to photon number is converged.

\begin{figure}[htbp]
    \centering
    \includegraphics[width=7.5cm]{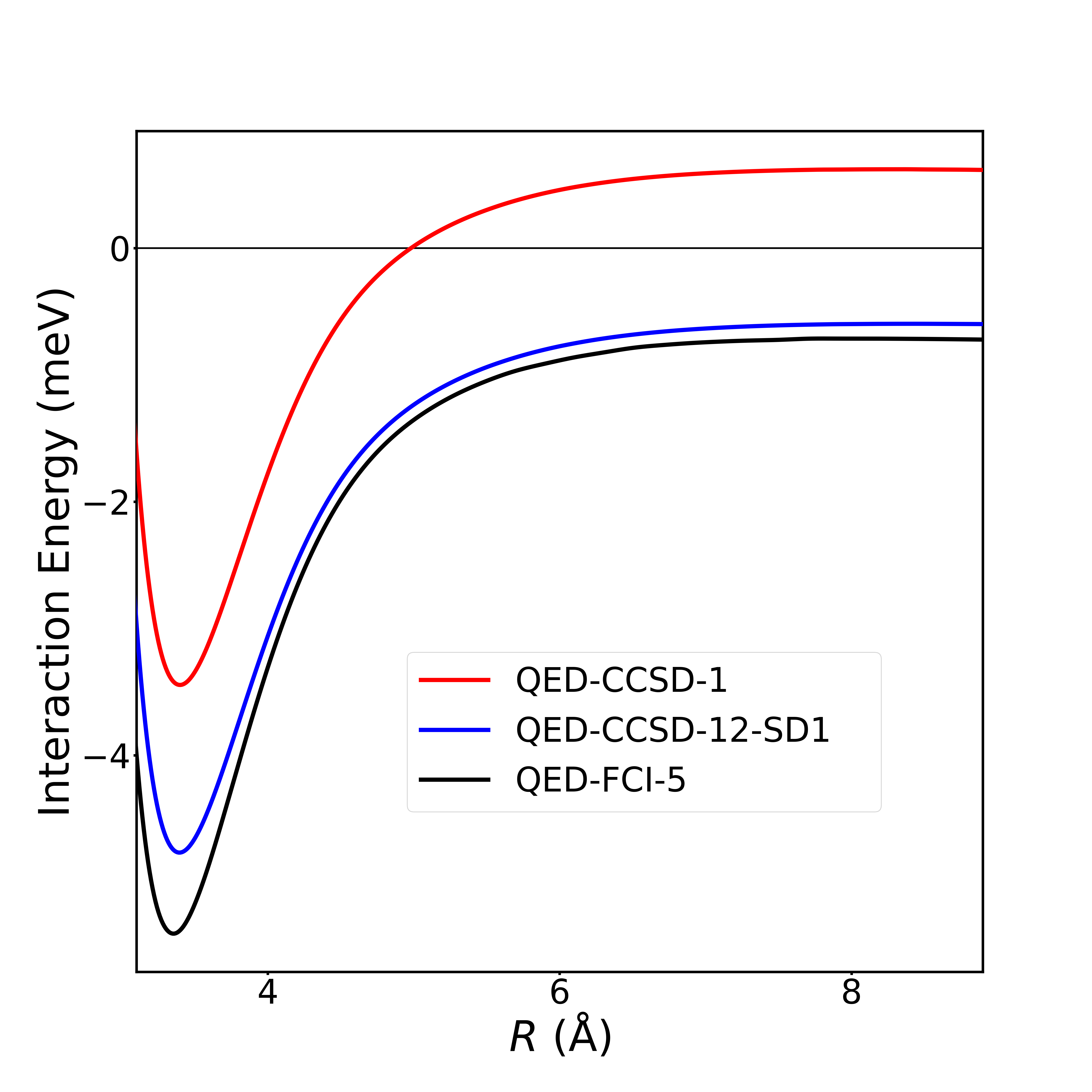}
    \caption{\label{fig:pes-qed-methods} Calculated intermolecular interaction energies for a C$_{\rm 2v}$ configuration of $2$H$_2$ with the cavity polarization vector parallel to the center-to-center intermolecular distance vector. All calculations shown in this figure were performed with $\lambda = 0.1$~a.u}
\end{figure}

We use QED-CCSD-12-SD1 instead of QED-CCSD-1 (equivalent to QED-CCSD-1-SD1) because the two-photon excitations are important for properly modeling the two-body interactions, as tested against QED-FCI-5 calculations. Without two-photon excitations, the two-body interactions have the wrong sign in the case of molecules separated by large distances (e.g. molecules separated by more than $1$~nm). This is visualized in \ref{fig:pes-qed-methods}.

In all of our calculations, we use a linearly polarized optical cavity with a single photon frequency and single polarization vector. In most experiments as of today, the optical cavity is not limited to just one polarization, but rather it hosts two degenerate cavity modes with orthogonal polarizations (both cavity mode polarization vectors are perpendicular to the cavity wavevector). Since the molecular orientations aligns with the transversal polarization, we expect that a standard optical cavity, which has both polarizations, will interact with the system differently. In particular, we expect that for few molecules, the molecules will orient along the wavevector $\vec{k}$, perpendicular to both cavity polarization vectors. For many molecules, we expect that the molecules will align perpendicular to $\vec{k}$, in the plane defined from the two transversal polarization vectors.

\section{Perturbation Theory}

As we demonstrate throughout this work, strong coupling to a single photon mode fundamentally changes the length scales and orientational dependence in which van der Waals molecules interact with one another. In this section, we explain these observations by performing perturbation theory in a similar spirit as Fritz London did in 1930\cite{Eisenschitz1930,London1930,London1937} but with additional perturbative potentials associated with coupling to the cavity. This analysis shows cavity-mediated intermolecular interactions between van der Waals molecules that scale with $R^{-3}$ and distance independent, $R^{0}$, interactions in addition to modifications to London dispersion forces that have an $R^{-6}$ dependence.\cite{Thirunamachandran1980,Milonni1996,Sherkunov2009,Fiscelli2020}

The total Hamiltonian is given by $H=H^0+H^1$ with
\begin{equation}
H^0 = H_{e,A} + H_{e,B} + \omega_c b^\dagger b
\label{eq:H^0}
\end{equation}
where $b^\dagger$ and $b$ are photon creation and annihilation operators for the cavity mode of frequency $\omega_c$ and $H_{e,A}$ and $H_{e,B}$ refer to the electronic Hamiltonians of molecules $A$ and $B$, respectively. The perturbative Hamiltonian ($H^1$) includes the dipolar coupling between molecules $A$ and $B$, in the spirit of London's first derivation of van der Waals interactions, and the light-matter coupling to a single cavity mode
\begin{widetext}
\begin{equation}
    H^{1} = -\frac{\vec{d}_A \cdot \vec{d}_B}{R^3} + \frac{3 (\vec{d}_A \cdot \vec{R})(\vec{d}_B \cdot \vec{R})}{R^5}
    + \lambda \sqrt{\frac{\omega_c}{2}} (\vec \epsilon \cdot \Delta \vec d_A + \vec \epsilon \cdot \Delta \vec d_B) (b + b^\dagger)
    + \frac{\lambda^2}{2} (\vec \epsilon \cdot \Delta \vec d_A + \vec\epsilon \cdot \Delta \vec d_B)^2
    \label{eq:H^1}
\end{equation}
\end{widetext}
where $\Delta \vec d_A = \vec d_A - \langle\vec d_A\rangle$ and $\Delta\vec d_B = \vec d_B - \langle\vec d_B\rangle$ are the fluctuations of molecule $A$ and molecule $B$'s dipoles, respectively and $\vec d_A$ and $\vec d_B$ are the dipole operators for molecule $A$ and molecule $B$, respectively. Recall that in this work we are working with van der Waals molecules such that both molecules do not have permanent dipoles (i.e. $\langle\vec d_A\rangle=\langle\vec d_B\rangle=0$).

The first-order correction to the energy is given by
\begin{equation}
E^1 = \matrixel{g}{H^1}{g}
\label{eq:E^1-definition}
\end{equation}
where $\ket{g}$ denotes the ground state of the total system, $\ket{g}=\ket{g_A}\ket{g_B}\ket{g_c}$ where molecule $A$, molecule $B$, and the cavity are in their ground states. In this illustrative perturbation theory, we are interested in the asymptotic behavior for when molecule $A$ and molecule $B$ are far away from one another; thus, the antisymmetry of the total electronic wavefunctions is ignored. Substituting in Eq.~\ref{eq:H^1} into Eq.~\ref{eq:E^1-definition}, we obtain
\begin{widetext}
\begin{align}
    E^1 & = \frac{\lambda^2}{2} (\matrixel{g_A}{\left( \vec{d_{A}} \cdot \vec{\epsilon} \right)^2}{g_A} + \matrixel{g_B}{\left( \vec{d_{B}} \cdot \vec{\epsilon} \right)^2}{g_B}) \nonumber \\
    & = \frac{\lambda^2}{2}(E^1_A+E^1_B)
    \label{eq:E^1}
\end{align}
\end{widetext}
where $E^1_A=\matrixel{g_A}{\left( \vec{d_{A}} \cdot \vec{\epsilon} \right)^2}{g_A}$ and $E^1_B=\matrixel{g_B}{\left( \vec{d_{B}} \cdot \vec{\epsilon} \right)^2}{g_B}$ are the dipole self-energies of molecule $A$ and molecule $B$, respectively. In Eq.~\ref{eq:E^1} we have used the facts that there are no photons in the ground state of the cavity ($\matrixel{g_c}{b^\dagger b}{g_c}=0$) and that for van der Waals molecules, by definition, there is no permanent dipole ($\matrixel{g_A}{\vec{d}_A}{g_A}=\langle\vec d_A\rangle=0$ and $\matrixel{g_B}{\vec{d}_B}{g_B}=\langle\vec d_B\rangle=0$). The fact that molecules $A$ and $B$ do not have permanent dipoles allows us to express $E^1_A$ and $E^1_B$ with a different formula, i.e. 
\begin{align}
    E^1_A & = \matrixel{g_A}{\left( \vec{d_{A}} \cdot \vec{\epsilon} \right)^2}{g_A} \\ \nonumber 
    & = \matrixel{g_A}{\left( \vec{d_{A}} \cdot \vec{\epsilon} \right)\hat{I}\left( \vec{d_{A}} \cdot \vec{\epsilon} \right)}{g_A} \\ \nonumber
    & = \sum_{e_A}|\matrixel{e_A}{\left( \vec{d_{A}} \cdot \vec{\epsilon} \right)}{g_A}|^2\;\;,
\end{align}
where $\ket{e_A}$ is an excited state of molecule $A$. An important observation here is that both $E^1_A$ and $E^1_B$ are single molecule terms and are always positive; we will return to these facts after deriving the second-order energy correction. 

The second-order correction to the energy is given by 
\begin{equation}
E^2 = -\sum_{e} \frac{\left| \matrixel{e}{H^1}{g} \right|^2}{E_{e} - E_{g}}
\label{eq:E^2-definition}
\end{equation}
where $\ket{g}$ is the ground state of the bi-molecule system with energy $E_g$ and $\ket{e}$ indicates an excited state of the bi-molecule system with energy $E_e$. Substituting Eq.~\ref{eq:H^1} into Eq.~\ref{eq:E^2-definition} along with some simplifications we obtain the second-order correction to the energy to be
\begin{widetext}
\begin{align}
\label{eq:E^2} E^2 = \nonumber & -\sum_{e_A e_B} \frac{\left| \matrixel{e_A e_B}{V_{AB}}{g_A g_B} \right|^2}{E_{e_A} - E_{g_A} + E_{e_B} - E_{g_B}} - \lambda^2 \sum_{e_A e_B} \frac{\matrixel{e_A e_B}{V_{AB}}{g_A g_B} \matrixel{e_A}{\vec{d_A} \cdot \vec{\epsilon}}{g_A} \matrixel{e_B}{\vec{d_B} \cdot \vec{\epsilon}}{g_B}}{E_{e_A} - E_{g_A} + E_{e_B} - E_{g_B}} \\ \nonumber & - \frac{\lambda^2 \omega_c}{2} \left[ \sum_{e_A} \frac{\left| \matrixel{e_A}{\vec{d_A} \cdot \vec{\epsilon}}{g_A} \right|^2}{\omega_c + E_{e_A} - E_{g_A}} + \sum_{e_B} \frac{\left| \matrixel{e_B}{\vec{d_B} \cdot \vec{\epsilon}}{g_B} \right|^2}{\omega_c + E_{e_B} - E_{g_B}} \right] \\ \nonumber & -\frac{\lambda^4}{4} \left[ \sum_{e_A} \frac{\left| \matrixel{e_A}{\left( \vec{d_{A}} \cdot \vec{\epsilon} \right)^2}{g_A} \right|^2}{E_{e_A} - E_{g_A}} + \sum_{e_B} \frac{\left| \matrixel{e_B}{\left( \vec{d_{B}} \cdot \vec{\epsilon} \right)^2}{g_B} \right|^2}{E_{e_B} - E_{g_B}} + 4\sum_{e_A e_B} \frac{\left| \matrixel{e_A}{\left( \vec{d_{A}} \cdot \vec{\epsilon} \right)}{g_A} \right|^2 \left| \matrixel{e_B}{\left( \vec{d_{B}} \cdot \vec{\epsilon} \right)}{g_B} \right|^2}{E_{e_A} - E_{g_A}+ E_{e_B} - E_{g_B}} \right] \\ 
& = E^2_{AB,d^0}+\lambda^2 E^2_{AB,d^1}+\frac{\lambda^2}{2}(E^2_{A,d^1}+E^2_{B,d^1})+\frac{\lambda^4}{4}(E^2_{A,d^2}+E^2_{B,d^2}+E^2_{AB,d^2})
\end{align}
\end{widetext}
where we defined
\begin{equation}
V_{AB} = -\frac{\vec{d}_A \cdot \vec{d}_B}{R^3} + \frac{3 (\vec{d}_A \cdot \vec{R})(\vec{d}_B \cdot \vec{R})}{R^5}\;\;\ldotp
\label{eq:V_AB-definition}
\end{equation}
$E^2_{AB,d^0}$, $E^2_{AB,d^1}$, $E^2_{A,d^1}$, $E^2_{B,d^1})$, $E^2_{A,d^2}$, $E^2_{B,d^2}$, and $E^2_{AB,d^2}$ are defined as 
\begin{widetext}
\begin{subequations}
\begin{align}
E^2_{AB,d^0} & = -\sum_{e_A e_B} \frac{\left| \matrixel{e_A e_B}{V_{AB}}{g_A g_B} \right|^2}{E_{e_A} - E_{g_A} + E_{e_B} - E_{g_B}} \\
E^2_{AB,d^1} & = -\sum_{e_A e_B} \frac{\matrixel{e_A e_B}{V_{AB}}{g_A g_B} \matrixel{e_A}{\vec{d_A} \cdot \vec{\epsilon}}{g_A} \matrixel{e_B}{\vec{d_B} \cdot \vec{\epsilon}}{g_B}}{E_{e_A} - E_{g_A} + E_{e_B} - E_{g_B}} \\
E^2_{A,d^1} & = -\omega_c\sum_{e_A} \frac{\left| \matrixel{e_A}{\vec{d_A} \cdot \vec{\epsilon}}{g_A} \right|^2}{\omega_c + E_{e_A} - E_{g_A}} \\
E^2_{B,d^1} & = -\omega_c\sum_{e_B} \frac{\left| \matrixel{e_B}{\vec{d_B} \cdot \vec{\epsilon}}{g_B} \right|^2}{\omega_c + E_{e_B} - E_{g_B}} \\
E^2_{A,d^2} & = -\sum_{e_A} \frac{\left| \matrixel{e_A}{\left( \vec{d_{A}} \cdot \vec{\epsilon} \right)^2}{g_A} \right|^2}{E_{e_A} - E_{g_A}} \\
E^2_{B,d^2} & = -\sum_{e_B} \frac{\left| \matrixel{e_B}{\left( \vec{d_{B}} \cdot \vec{\epsilon} \right)^2}{g_B} \right|^2}{E_{e_B} - E_{g_B}} \\
E^2_{AB,d^2} & = -4\sum_{e_A e_B} \frac{\left| \matrixel{e_A}{\left( \vec{d_{A}} \cdot \vec{\epsilon} \right)}{g_A} \right|^2 \left| \matrixel{e_B}{\left( \vec{d_{B}} \cdot \vec{\epsilon} \right)}{g_B} \right|^2}{E_{e_A} - E_{g_A}+ E_{e_B} - E_{g_B}} \;\;,
\end{align}
\end{subequations}
\end{widetext}
where $\ket{g_A}$ ($\ket{g_B}$) is the ground state of molecule $A$ ($B$) with energy $E_{g_A}$ ($E_{g_B}$), $\ket{e_A}$ ($\ket{e_B}$) indicates an excited state of molecule $A$ ($B$) with energy $E_{e_A}$ ($E_{e_B}$), and $\matrixel{e_A}{\vec{d}_A}{g_A}$ ($\matrixel{e_B}{\vec{d}_B}{g_B}$) is the transition dipole moment of molecule $A$ ($B$) associated with the excited state.
Eq.~\ref{eq:E^2} is an important result in this work, and the physical interpretation, origin, and implications of each term are worth exploring in detail. $E^2_{AB,d^0}$ in Eq.~\ref{eq:E^2} is the typical attractive London dispersion interaction with its prototypical $R^{-6}$ dependence (as each $V_{AB}$ scales with $R^{-3}$). The remaining terms all arise from interactions through the cavity mode. $E^2_{AB,d^1}$ contains a single $V_{AB}$ matrix element giving an $R^{-3}$ of this term. Interestingly, this term also contains dot products of transition dipole moments ($\matrixel{e_A}{\vec{d}_A}{g_A}$) with the cavity polarization vector ($\vec{\epsilon}$). This $R^{-3}$ term is central to this work as it says that van der Waals molecules inside a cavity have this interesting interaction length scale that also has unique, coupled molecule-molecule and molecular-cavity angle dependencies. $E^2_{A,d^1}$ and $E^2_{B,d^1}$are very similar to $E^1_A$ and $E^1_B$ except that $E^2_{A,d^1}$ and $E^2_{B,d^1}$ arise from the bilinear coupling term and have the opposite sign as $E^1_A$ and $E^1_B$. Specifically, to second-order in the coupling $\lambda$, the one-body energy (e.g. molecule $A$) is given by
\begin{align}
    \label{eq:1-body-energy-lambda-squared}
    E_{A}^{\text{cavity}} & = E_{A}^{\text{no cavity}}+\frac{\lambda^2}{2}(E^1_A+E^2_{A,d^1}) \\ \nonumber 
    & = E_{A}^{\text{no cavity}} + \frac{\lambda^2}{2} \sum_{e_A} \left| \matrixel{e_A}{\vec{d}_A \cdot \vec{\epsilon}}{g_A} \right|^2  \\ \nonumber
    &\quad- \frac{\lambda^2 \omega_c}{2} \sum_{e_A} \frac{\left| \matrixel{e_A}{\vec{d}_A \cdot \vec{\epsilon}}{g_A} \right|^2}{w_c + E_{e_A} - E_{g_A}}\;\;\ldotp
\end{align}
A similar energy term can be derived for molecule $B$ as well. 
We want to emphasize that $E^1_A$ arises from the dipole self-energy term in first-order perturbation theory (Eq.~\ref{eq:E^1}) and $E^2_{A,d^1}$ arises from the bilinear coupling term in second-order perturbation theory (Eq.~\ref{eq:E^2}). Interestingly, $E^1_A$ and $E^2_{A,d^1}$ only exactly cancel if the cavity frequency is much larger than the electronic transition energies ($\omega_c \gg E_{e_A}-E_{g_A}$). Thus, for H$_2$ molecules with a cavity in the electronic regime ($\omega_c = 13.6$~eV here) the total energy of a single molecule ends up increasing with $\lambda^2$ (main text Fig.~3A). For H$_2$ molecules, the one-body energy reaches a minimum when the molecular bond is perpendicular to the cavity polarization vector ($\theta_{A\epsilon}=\frac{\pi}{2}$). Intuitively, this occurs because H$_2$ is most polarizable along its bond axis which leads to $\sum_{e_A} \left| \matrixel{e_A}{\vec{d}_A \cdot \vec{\epsilon}}{g_A} \right|^2 / (E_{e_A} - E_{g_A}) = \vec \epsilon^T \bm{\alpha} \vec \epsilon$ being largest when $\theta_{A\epsilon}=0,\pi$.

$E^2_{A,d^2}$, $E^2_{B,d^2}$, and $E^2_{AB,d^2}$ arise from two factors of the dipole self-energy part of Eq.~\ref{eq:H^1} and, thus, scale with $\lambda^4$. While $E^2_{A,d^2}$ and $E^2_{B,d^2}$ are corrections to the one-body energies, $E^2_{AB,d^2}$ impacts the two-body energies (i.e. intermolecular interaction energy). Furthermore, this term has no $R$ dependence, and, thus, $E^2_{AB,d^2}$ is the first term that we have discussed that gives rise to the collective orientational order reported in the main text. The magnitude of this term is greatest when both molecules have their bonds oriented along the cavity polarization vector ($\vec{\epsilon}$), because $\vec \epsilon^T \bm{\alpha}_A \vec \epsilon$ and $\vec \epsilon^T \bm{\alpha}_B \vec \epsilon$ are both largest in the case which both of their bonds are oriented parallel to $\vec{\epsilon}$. And because of the negative sign in front of this infinite range interaction term, it contributes to lowering the energy of molecular configurations in which the molecular bonds of the hydrogen molecules are oriented parallel to the cavity polarization vector, as shown in Fig.~3C of the main text.


\section{Many-body Interactions}

\begin{figure}[htb]
    \centering
    \includegraphics[width=\linewidth]{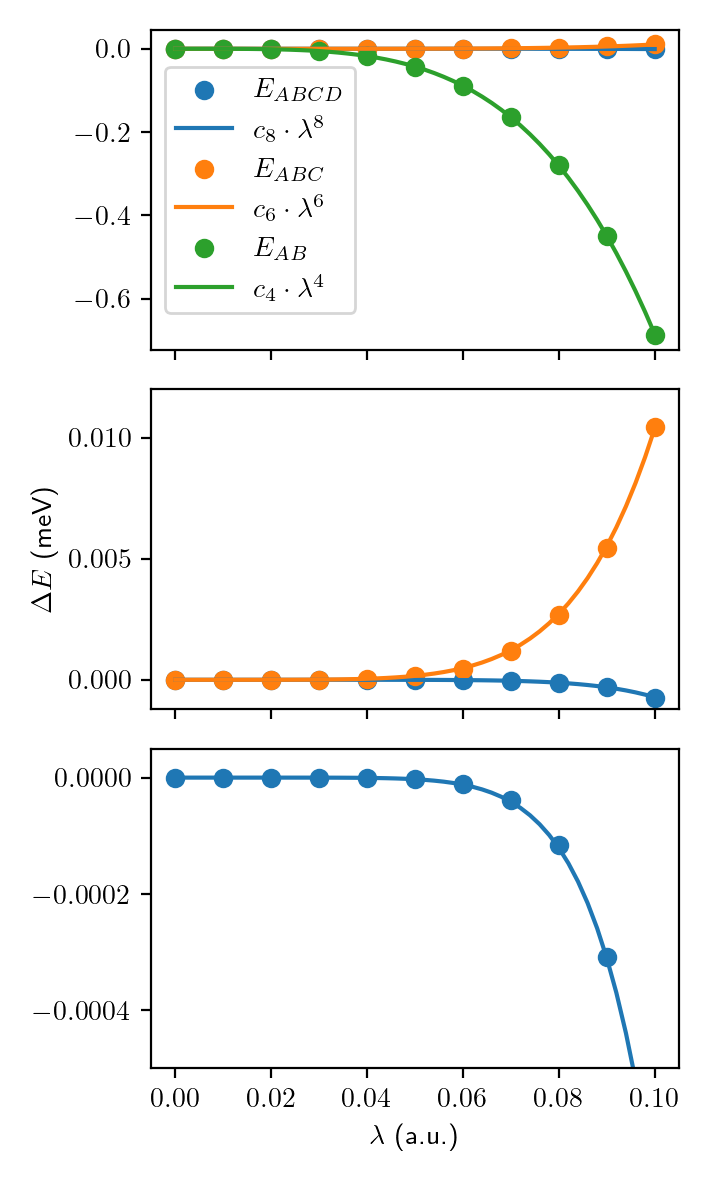}
    \caption{$N$-body effects for different coupling strengths $\lambda$. All calculations are performed on $N$ H$_2$ molecules with QED-CCSD-12-SD1. The cavity energy is $\hbar \omega_c = 13.6$~eV and polarization perpendicular to all molecules. The molecules are placed on the edges of a line ($E_{AB}$), equilateral triangle ($E_{ABC}$) and square ($E_{ABCD}$), all with side lengths of $25$~Å.}
    \label{fig:lambda_E_nbody}
\end{figure}

The many-body expansion,
\begin{equation}
    E = \sum_A E_A + \sum_{AB} E_{AB} + \sum_{ABC} E_{ABC} + \dots
\end{equation}
is a routinely used expansion for modeling and gaining insight into intermolecular forces.\cite{Dahlke2007} For van der Waals type intermolecular forces, the higher-order interactions such as $E_{ABC}$ quickly become negligible with distance and they can be assumed to be much smaller than the lower-order terms at large distances. QED electronic structure calculations allow us to test if the three-body and higher-order terms can be ignored for the strong light-matter coupling cavity QED Hamiltonian with similar parameters used in the calculations of the main text. In \ref{tab:nbody_numbers} and \ref{fig:lambda_E_nbody}, we show the intermolecular interactions for molecules separated far apart, $25$~Å. As expected, QED-HF does not capture the dynamic correlation and cannot describe the intermolecular forces arising from neither the cavity nor the van der Waals forces. QED-CCSD-1 captures the dynamic correlation, but the sign of the two-body interaction is not consistent with QED-FCI. Adding just one more term to the cluster operator of QED-CCSD-1, the two-photon $(b^\dagger)^2$ term in QED-CCSD-12-SD1, yields a sufficient description of the two-body interactions. For QED-CCSD-12-SD1, we find that the higher-order terms quickly approach zero even for the very strong coupling $\lambda = 0.1$~a.u. From perturbation theory, we find that the $N$-body interactions are sensitive to the light-matter coupling strength and scale as $\lambda^{2N}$ (see \ref{fig:lambda_E_nbody}).

A few additional key points about the many-body expansion of van der Waals interactions in the context of the nonrelativistic cavity QED Hamiltonian given in Eq.~\ref{eq:H_total} are worth mentioning here. Because the three-body interactions have opposite sign to the two-body interactions (\ref{tab:nbody_numbers}), we expect that the collective orientational order induced by the infinite range cavity-induced interactions would be reduced by including the three-body terms in the molecular dynamics simulations. While the three-body terms are insignificant on a per interaction basis, the lack of distance ($R$) dependence in the cavity-induced interactions, see Eq.~\ref{eq:E^2}, results in all molecules in the simulation interacting with all other molecules independent of how far away they are from each other. In a simulation with $n$ molecules, there are $n(n-1)/2\sim n^2$ two-body interactions, $n(n-1)(n-2)/6 \sim n^3$ three-body interactions, and similarly for higher-order terms (\ref{tab:nbody_scaling}). Therefore, there must exist a number of molecules where the total three-body energy is larger than the total two-body energy. This makes it very challenging to extrapolate our results to truly macroscopic systems. Extending these microscopic equations and calculations to truly macroscopic systems remains an open question.

\begin{table}[htb]
    \centering
    \begin{tabular}{lcccc}
        \hline
        Method  & 1-body & 2-body & 3-body & 4-body \\
        \hline
        QED-HF          & 204.9 &  0.0000 &  0.0000  & 0.0000 \\
        QED-CCSD-1      & 107.5 &  0.3238 & -0.0571 &  0.0042 \\
        QED-CCSD-12-SD1 & 107.1 & -0.5600 &  0.0104 & -0.0004 \\
        QED-FCI-5       & 106.7 & -0.6601 & $\ldots$ & $\ldots$ \\
        \hline
    \end{tabular}
    \caption{Cavity-induced $N$-body effects for different QED electronic structure methodologies with $\lambda=0.1$ a.u. The cavity energy is $\hbar \omega_c = 13.6$~eV and polarization perpendicular to all molecules. The molecules are placed on the edges of a line ($E_{AB}$), equilateral triangle ($E_{ABC}$) and square ($E_{ABCD}$), all with side lengths of $25$~Å. All numbers in the table are meV. QED-FCI-5 is too computationally expensive for more than two H$_2$ molecules in the aug-cc-pVDZ basis set.}
    \label{tab:nbody_numbers}
\end{table}

\begin{table}[htb]
    \centering
    \begin{tabular}{lcccc}
        \hline
         & 1-body & 2-body & 3-body & 4-body \\
        \hline
        Scaling with coupling & $\lambda^2$ &  $\lambda^4$   & $\lambda^6$     & $\lambda^8$  \\
        Number of terms & $n \choose 1$ & $n \choose 2$ & $n \choose 3$ & $n \choose 4$ \\
        \hline
    \end{tabular}
    \caption{The number of interactions and scaling of the cavity-induced interaction energy in the $N$th body of the $N$-body expansion for a system with $n$ molecules.}
    \label{tab:nbody_scaling}
\end{table}

\section{Molecular Dynamics}

\subsection{Training Potential Energy Functions for Simulating Fluids of H$_2$}

\subsubsection{Neural Network-based Pairwise Interactions}

\begin{figure}
    \centering
    \includegraphics[width=8.0cm]{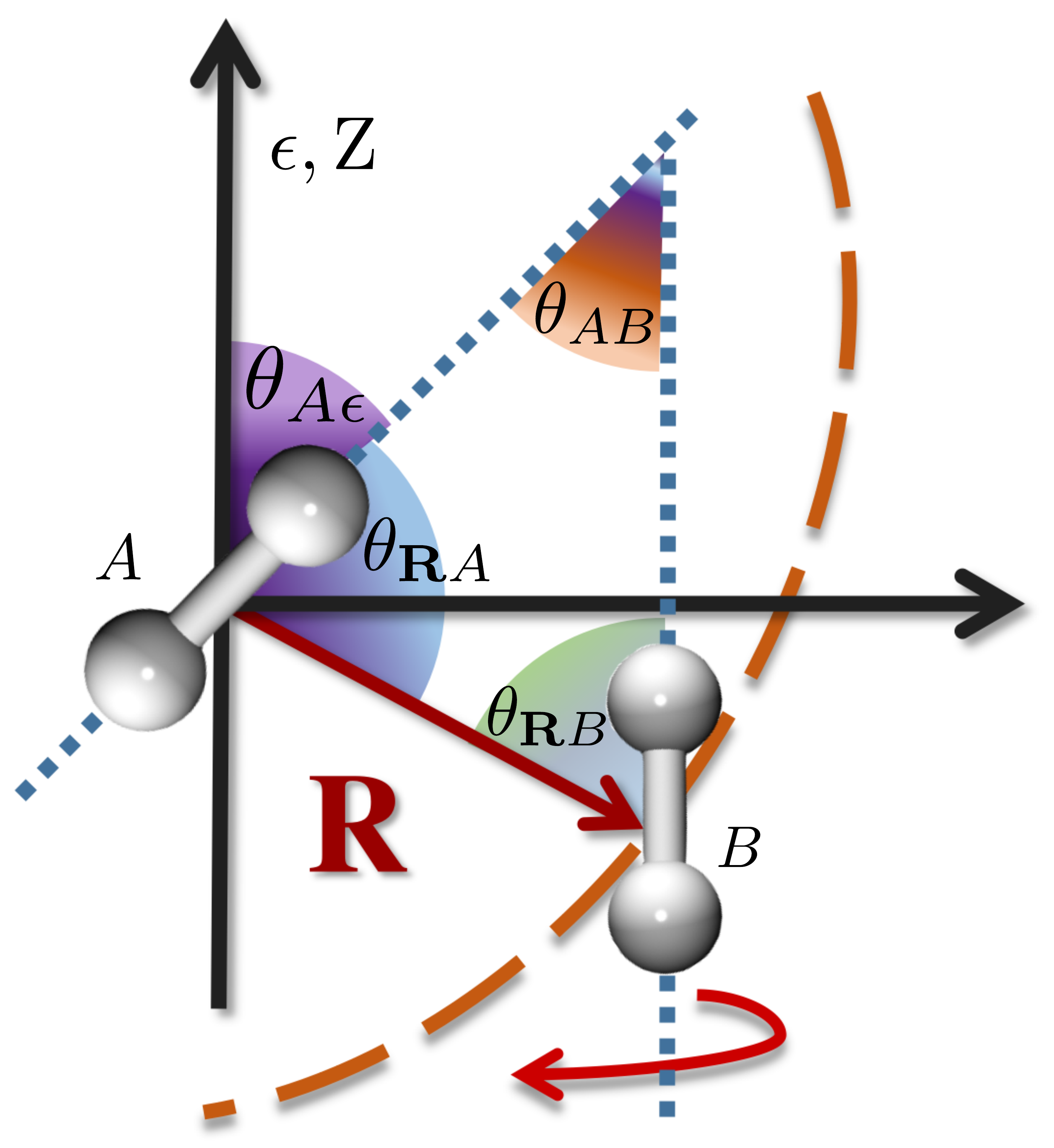}
    \caption{Symmetry preserved features that are considered while generating the pair interaction potential using a neural network based machine learning model are shown here. Various angles between a pair of molecule which are considered as input features are shown. $\bf R$ is the distance vector of the center of mass (COM) of molecule $A$ and molecule $B$. $\epsilon$ represents the cavity polarization. Orientation of the molecules are completely specified by various angles $\{\theta\}$.}
    \label{fig:features}
\end{figure}

We developed neural network-based potential energy functions (NNPs) for the pairwise interaction of a pair of hydrogen molecules using ${\it ab~initio}$ energy data with CCSD, FCI, QED-CCSD-12-SD1, and QED-FCI levels of theory.  The potential energy functions have the forms,
\begin{equation}
    E_{\rm AB}^{\rm no~cavity} = c_{\rm exp}\exp(-aR) - \frac{c_{6}\{\theta\}}{R^{6}} 
\end{equation}
\begin{equation}
    E_{\rm AB}^{\rm cavity} = E_{\rm 2b}^{\rm no~cavity} - \frac{c_{3}\{\theta\}}{R^{3}} + \frac{c_{0}\{\theta\}}{R^{0}} 
\end{equation}
where $c_{\rm exp}, {a}, c_6, c_3, c_0$ are represented by neural networks (NNs). Each NN takes symmetry preserved features of a pair of molecules as input.  Symmetry preserved features that have been selected as the input for the machine learning (ML) model to get the pairwise interaction energy are shown pictorially in~\ref{fig:features} and are listed in~\ref{tab:features}. In the case without the cavity field, the interaction energies are obtained using the input features $\theta_{{\bf R}A}, \theta_{{\bf R}B}, \theta_{AB}, {\left \| \bf R \right \|}$.  
For the cavity case, additional terms that depend on the cavity polarization vector are added. In particular, $\theta_{A\epsilon}, \theta_{B\epsilon}, \text{ and } \theta_{{\bf R}\epsilon}$ are added and $\left \| \bf R \right \|$ is replaced by  $R_{\rm cap}$ and $R_{\rm cap}= {\rm C} \tanh(\left \| \bf R \right \|/{\rm C})$, where C is a cutoff distance.  In order to account for molecular and exchange symmetries, $\cos2\theta$ and $\sin2\theta$ are used for any $\theta \in \Theta\equiv\{ \theta_{{\bf R}A},\theta_{{\bf R}B},\theta_{\rm AB}, \theta_{A\epsilon},\theta_{B\epsilon}, \theta_{{\bf R}\epsilon}\}$. For each of $c_{\rm exp}$, $a$, $c_6$, $c_3$, we are using F($\Theta,R_{\rm cap}$)+F($\tilde{\Theta},\tilde{R}_{\rm cap}$) where $\tilde{\Theta}$ and $\tilde{R}_{\rm cap}$ are calculated by switching the index of the two molecules. For $c_0$, only Type~1 features as tabulated in~\ref{tab:features} were used.

The neural network model has four fully-connected layers including a linear output layer. The other three linear layers have CELU activation functions.\cite{CELU_paper} The number of neurons per layer is 64 in our model. To train the model, we used energy data points of pair configurations that are generated using a classical MD simulation of liquid H$_2$. $10^5$ pair configurations generated by MD simulation were used to compute energies with CCSD level of theory for training model when no cavity is present. While the pair configurations generated by MD simulation were good enough to train a model without a cavity, long range pair configurations are extremely important to train the model with a cavity. Similarly, short range pair configurations are very crucial to accurately reproduce the corrected short range repulsion energies in the potential energy functions in the presence of a cavity. While MD of liquid H$_2$ produces good random configurations with various possible orientations, the probability of finding short range pair configurations is low in an MD simulation. In order to include sufficient number of configurations at short range, we randomly select $10\%$ of the total configurations obtained from MD simulation of liquid H$_2$ molecules and scale the intermolecular distance to be within $2-5$~{\AA}. A similar strategy was followed to generate very long range configurations between $18-90$~{\AA} for $10\%$ of the total configurations. A total of $121,000$ data points, including both the additional short range and long range configurations, were used to the train the NN model to the QED-CCSD-12-SD1 calculated energies in the cavity case. For training using the QED-FCI calculated data, we use a smaller data set of $30,000$ calculated energies. In order to train the model on this smaller data set, we initialize each NN with the parameters obtained from our QED-CCSD-12-SD1 fits, which was trained using a larger data set of $121,000$ calculated energies.
We use the Adam optimizer~\cite{Adam_optimizer} with $\beta_1 = 0.90$ and $\beta_2 = 0.99$. And we utilize a constant learning rate of $10^{-5}$ and a batch size of $32$. $90\%$ of the total data points were used in the training data set and the remaining $10\%$ were used as a test data set. All training and testing protocols were implemented with PyTorch.\cite{NEURIPS2019_9015}

\begin{table}
\begin{tabular}{ |c|c| } 
 \hline
 Type of feature & Features \\ \hline
 Type 1 & $\cos2\theta_{A\epsilon}$, $\sin2\theta_{A\epsilon}$,$\cos2\theta_{B\epsilon}$, $\sin2\theta_{B\epsilon}$  \\ \hline
 Type 2 & $\cos2\theta_{\bf R\epsilon}$, $\sin2\theta_{\bf R\epsilon}$,  $\cos2\theta_{{\bf R}A}$, $\sin2\theta_{{\bf R}A}$ \\ 
        & $\cos2\theta_{{\bf R}B}$, $\sin2\theta_{{\bf R}B}$, $\cos2\theta_{{AB}}$, $\sin2\theta_{{AB}}$ \\ \hline
 Type 3 & $\rm C \tanh(\left \| \bf R \right \|/C)$  \\ 
 \hline
\end{tabular}
\caption{\label{tab:features}Input features involved in the energy contributions.}
\end{table}

\begin{figure}
    \centering
    \includegraphics[angle=-90,width=8.0cm]{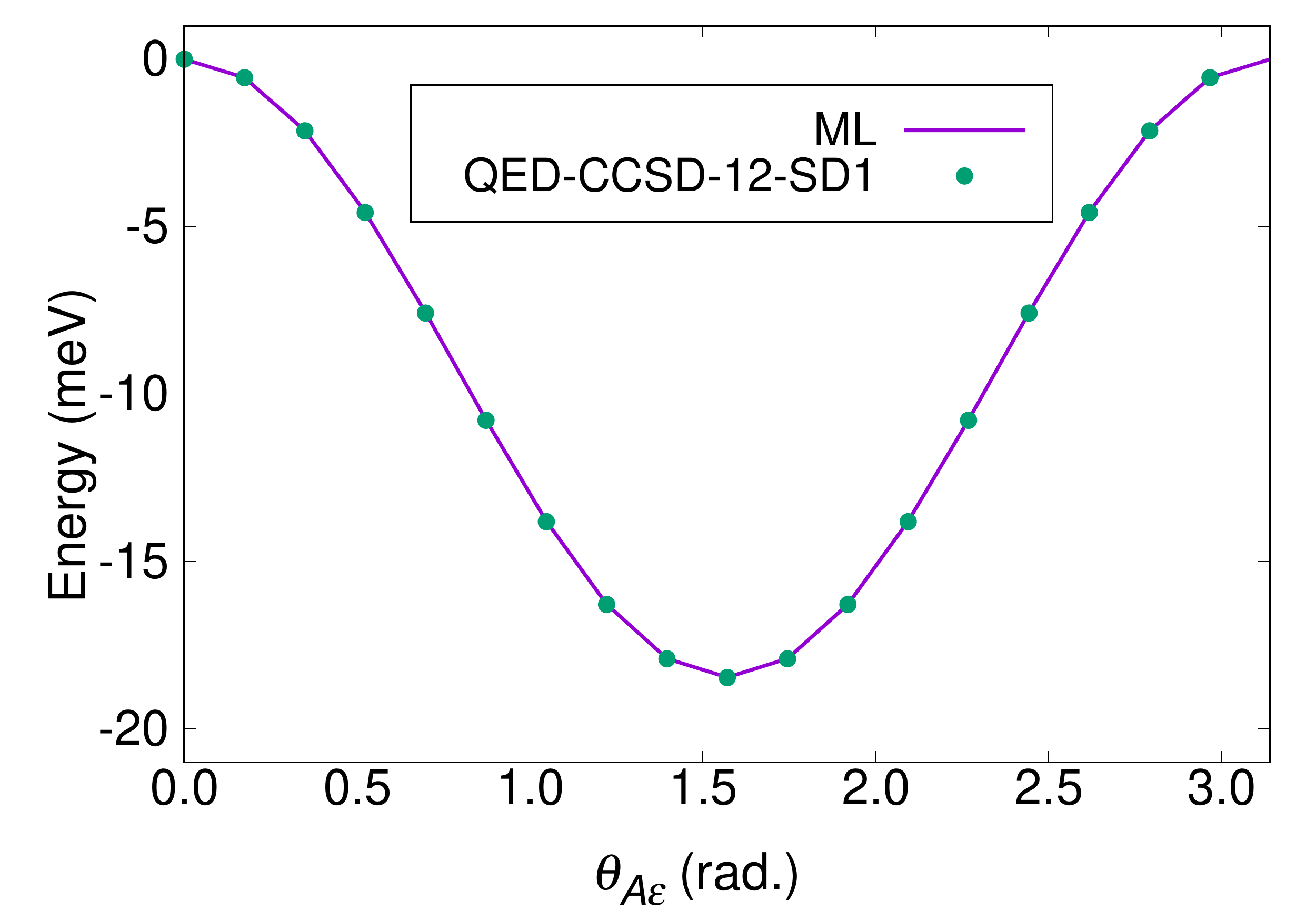}
    \caption{Energy of a single H$_2$ molecule inside a cavity with respect to cavity polarization vector, ${\epsilon}$ using {\ it ab initio} QED-CCSD-12-SD1 and ML. Single molecular energy at ${\epsilon} =0.0$ was set to zero while plotting energies of both QED-CCSD-12-SD1 and ML.}
    \label{fig:theta_1benergy}
\end{figure}

The energies of the {\it ab initio} (CCSD) calculations and the ML predicted energies of the pairs of molecules without a cavity field are shown in the~\ref{fig:nocavity_pes}A. A linearity plot shows the accuracy of the predicted energy using our ML model. Apart from the linearity plot, we scanned potential energy curves for a few selected orientations of pairs of molecules. These results show that the ML predicted potential energy curves for pairs of hydrogen molecules are in good agreement with the potential energy curves obtained from {\it ab initio} calculations. These plots are shown in~\ref{fig:nocavity_pes}B. 
A linearity plot comparing the {\it ab initio} (QED-CCSD-12-SD1) calculations and the ML predicted energies with the cavity field turned on are shown in~\ref{fig:cavity_pes}A. Potential energy curves (\ref{fig:cavity_pes}B) were scanned for D$_{\rm 2h}$ configuration of a pair of molecules along three different cavity polarization directions with respect to the molecular bond axis. These plots shows that our ML model accurately reproduces the {\it ab initio} potential energy curves. 

\subsubsection{Single Molecule Potential Energies}

Single molecule potential energies involve intra-molecular chemical bonds and the cavity-modified single molecule contributions. Intra-molecular chemical bonds were modeled within the harmonic approximation. We like to emphasize that the intra-molecular interaction energy does not play a significant role in determining the properties that we focused on in this study.

Single molecule energies in the presence of a cavity field is important. Training of the cavity-modified single molecule energies has been done with a linear regression method. The following form of energy function is trained for the single molecule energies,
\begin{equation}
E_{\rm A} =\sum_{n=1}^{3} C_{n}\sin2n{\theta} +\sum_{n=0}^{2} D_{n} \cos2n{\theta} 
\end{equation}
where $\theta$ is the angle between the molecular bond axis and the cavity polarization vector. $C_n$ and $D_n$ are the  trainable parameters. \ref{fig:theta_1benergy} shows the accuracy of fitting single molecular energies with respect to the {\it ab initio}, QED-CCSD-12-SD1 calculations. 

\subsection{Molecular Dynamics}

Molecular dynamics (MD) simulations were used to compute the statistical properties of fluids of $H_2$ molecules at $70$~K by employing the potential energy functions, generated by our machine learning models. For computing the statistical behaviour of the system both classical MD and path integral MD (PIMD) were used.

\subsubsection{Classical Molecular Dynamics}

NVT ensemble MD simulations were carried out using Langevin dynamics with a time step of $1.0$ femtosecond (fs) and the friction coefficient for the Langevin dynamics was chosen $0.0005$~a.u (20.7 ps$^{-1}$). Random initial atomic velocities and random initial positions were provided to run MD. In order to use ML potentials generated with PyTorch, we also implement the MD engine with PyTorch. The integrator used here is described in Ref.~\cite{Bussi_Langevin}. Forces were computed using the PyTorch autograd module and the PyTorch MD simulations were performed using GPUs. 

\begin{figure}
    \centering
    \includegraphics[width=7.0cm]{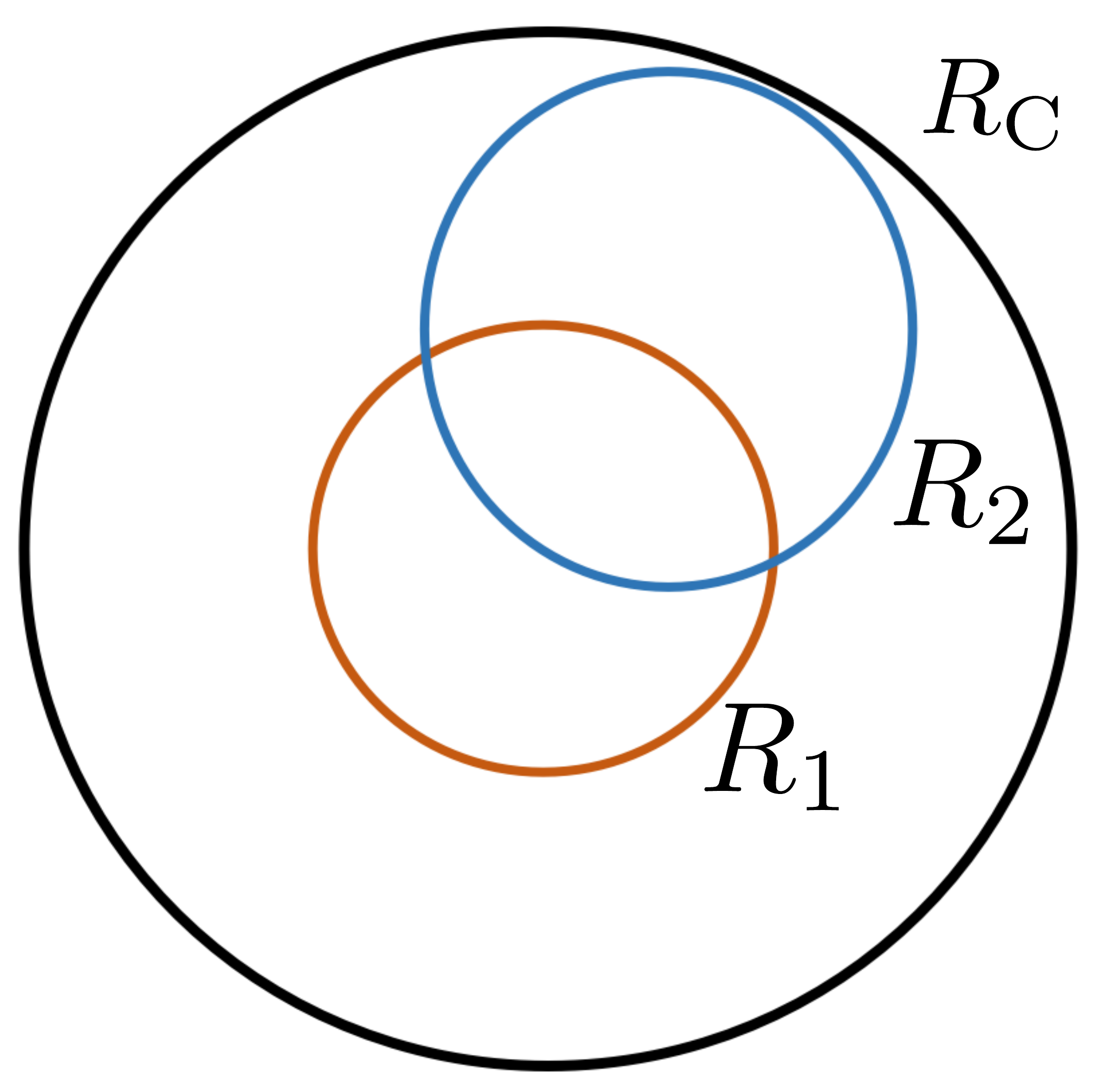}
    \caption{Schematic diagram of the radius cutoff that are used in computing radial distribution functions. $R_{\rm c}$ is the distance at which a high energy potential barrier has been applied. $R_{\rm 1}$ is the radius of core region where surface effects due to the spherical boundary are minimal and molecules found within the radius of $R_{\rm 2}$ are used to compute the histogram of pairwise distance for the calculations of the radial distribution functions.}
    \label{fig:boundary_RDF}
\end{figure}

Since we are simulating a fluid system, the system was confined within a spherical volume, similar to a cluster of molecules. In practice, a stiff harmonic potential was used to confine the center of the mass of each molecule within a spherical volume with radius $R_{\rm c}$ (see~\ref{fig:boundary_RDF}). Adopting such a boundary condition  was necessary in order to account for non-decaying nature of the pair interaction potential inside of an optical cavity. In order to simulate various different system sizes, $R_{\rm c}$ is scaled appropriately to preserve the overall molecular density.

\subsubsection{Path Integral Molecular Dynamics}

In the previous section, we discussed the MD simulations in which the nuclei were considered as classical particles. However, for light nuclei such as hydrogen atoms, this assumption could lead to serious problems in predicting the statistical properties because of strong quantum nuclei effects, especially at low temperatures. In order to account for quantum nuclei effects in our MD simulations, we performed path integral molecular dynamics (PIMD) simulations.

Usually PIMD simulations require a large number of beads to converge thermodynamics properties at low temperatures. Herein, we used the generalized Langevin equation (GLE) in PIMD, which can significantly reduce the number of beads.\cite{Ceriotti_PRL:2009,Ceriotti_GLE_2010,Ceriotti_PI-GLE_2011} In the GLE formulation,\cite{Ceriotti_JCP_2010} each bead of the simulated system is coupled to several extended degrees of freedom with an appropriate drift matrix and a diffusion matrix to approximate a friction kernel function. We used $8$ extra degrees of freedom in GLE and the drift matrix and diffusion matrix used in GLE were generated by an online tool called GLE4MD (\href{http://gle4md.org/}{http://gle4md.org/}) with the maximum physical frequency set to $\omega_{\rm max}=9608$~cm$^{-1}$. With the GLE formulation, we observed that using $32$ beads are able to converge the simulations whereas more than $128$ beads are needed to converge the results without the GLE formulation. We have developed an interface to i-PI~\cite{CERIOTTI20141019} to run the PIMD simulations using our ML potentials.

\subsection{Radial Distribution Functions}

The radial distribution functions (g(r)) of fluid of H$_2$ molecules are computed from the PIMD trajectories of $1,000$ molecules. As the system we simulated has a spherical volume without any periodic boundary, computing a bulk-like g(r) (i.e. a g(r) that converges to $1$ in the long distance limit) is not straightforward. In order to compute g(r) from such a spherical system, the following steps are taken. First, a bulk-like core region is chosen within a certain cutoff distance $R_1$.
\begin{equation}
 \bar h (\left | \bf r \right |)= \frac{1}{N_1}\sum_{i, R_i<R_1}^{} h(\left | {\bf r}-{\bf r}_i\right |)
\end{equation}
For the $i^{\rm th}$ molecule located at ${\bf r}_i$ with $R_i=\left |{\bf r}_i\right |<R_1$, $h(\left | {\bf r}-{\bf r}_i\right |)$ is the histogram of all distance between any other molecules and the $i^{th}$ molecule with $(\left | {\bf r}-{\bf r}_i\right |)< R_2$, $R_1 + R_2 <R_{\rm c}$ and $N_1$ is the number of molecules inside $R_1$. 
Second, the average over each frame of MD or PIMD as well as the average over number of beads was computed in the calculations of the radial distribution functions. Lastly, the averaged $\bar h (\left | \bf r \right |)$ was normalized by the average density and $4\pi r^2$. In this study, $R_1 = 6.0$~{\AA} and $R_2 = 12$~{\AA} was used. 

\subsection{Angular Distribution Functions}

We also computed angular distribution functions for the angle between the molecular bond axis of molecule $A$ and the molecular bond axis of molecule $B$ ($\theta_{\rm AB}$) and angular distribution functions for the angle between the molecular bond axis of molecule $A$ and the cavity polarization vector ($\theta_{\rm A\epsilon}$). The probability distributions of $\theta_{\rm AB}$ and $\theta_{\rm A\epsilon}$ are proportional to sin($\theta_{\rm AB}$) and sin($\theta_{\rm A\epsilon}$), respectively, if molecules A and B can rotate freely without any interactions. In order to emphasize the energy contribution, we computed the potentials of mean force by scaling the probability distributions of $\theta_{\rm AB}$ and $\theta_{\rm A\epsilon}$ with their corresponding sine functions. In the case of PIMD, the average over each frame and the average over the number of beads are considered when computing the histograms. 

\section{Additional Results}

\subsection{Comparison of Radial Distribution Functions}

We compute the radial distribution function at three different situations when (1) cavity polarization is not active, (2) cavity-modified one-body term is active but cavity modified two-body term is not active, and (3) both cavity modified one-body and two-body terms are active. We have observed differentiable changes in radial distribution function for three different situations. This indicates the difference in equilibrium structure when cavity polarization is on. The results are shown in~\ref{fig:RDF}. 

\subsection{Comparison of Classical MD and PIMD}

\begin{figure*}
    \centering
    \includegraphics[width=17.0cm]{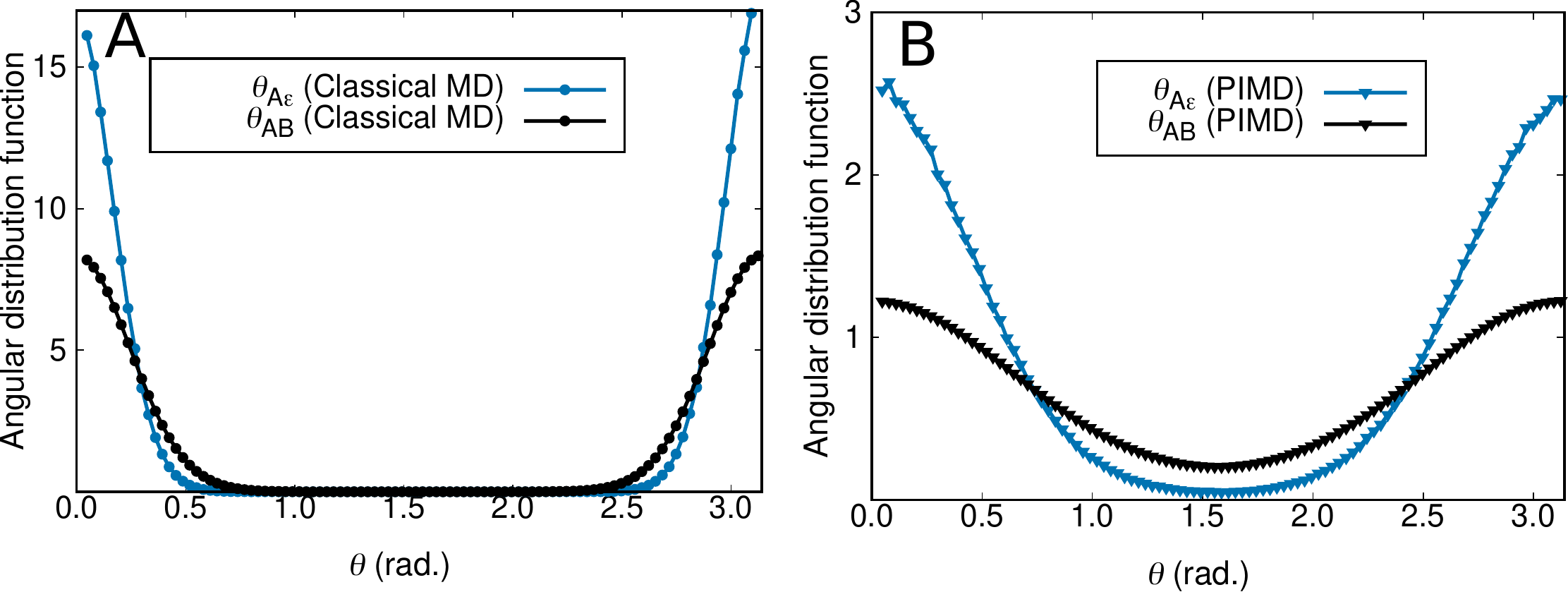}
    \caption{Angular distribution functions of molecular bond axis of molecule $A$ to the molecular bond axis of molecule $B$ ($\theta_{\rm AB}$) and angular distribution functions of molecular bond axis of molecule $A$ to the cavity polarization vector ($\theta_{\rm A\epsilon}$) for $1,000$ H$_2$ molecules of a (A) classical MD simulation and (B) PIMD simulation are shown. Pair interaction potentials used for the MD simulation were obtained by training an ML model with the calculated energies from QED-CCSD-12-SD1 level of theory.}
    \label{fig:classicalVsPIMD}
\end{figure*}

In this section we compare the results of our classical MD and the PIMD simulations with $\lambda=0.1$~a.u. Based on~\ref{fig:classicalVsPIMD}, it is evident that classical MD and PIMD qualitatively follow the same trend when angular distribution function of $\theta_{\rm A\epsilon}$ and $\theta_{\rm AB}$ are compared. In particular, one observes a strong orientational alignment of the molecules along direction of the cavity polarization vector occurring inside of an optical cavity. Inclusion of nuclear quantum effects does not change the overall conclusion. However, the extent of alignment of the molecules inside the cavity in our PIMD simulations is considerably reduced compared to our classical MD simulations.

\subsection{Comparison of QED-FCI-5 and QED-CCSD-12-SD1}

\begin{figure*}
    \centering
    \includegraphics[width=17.0cm]{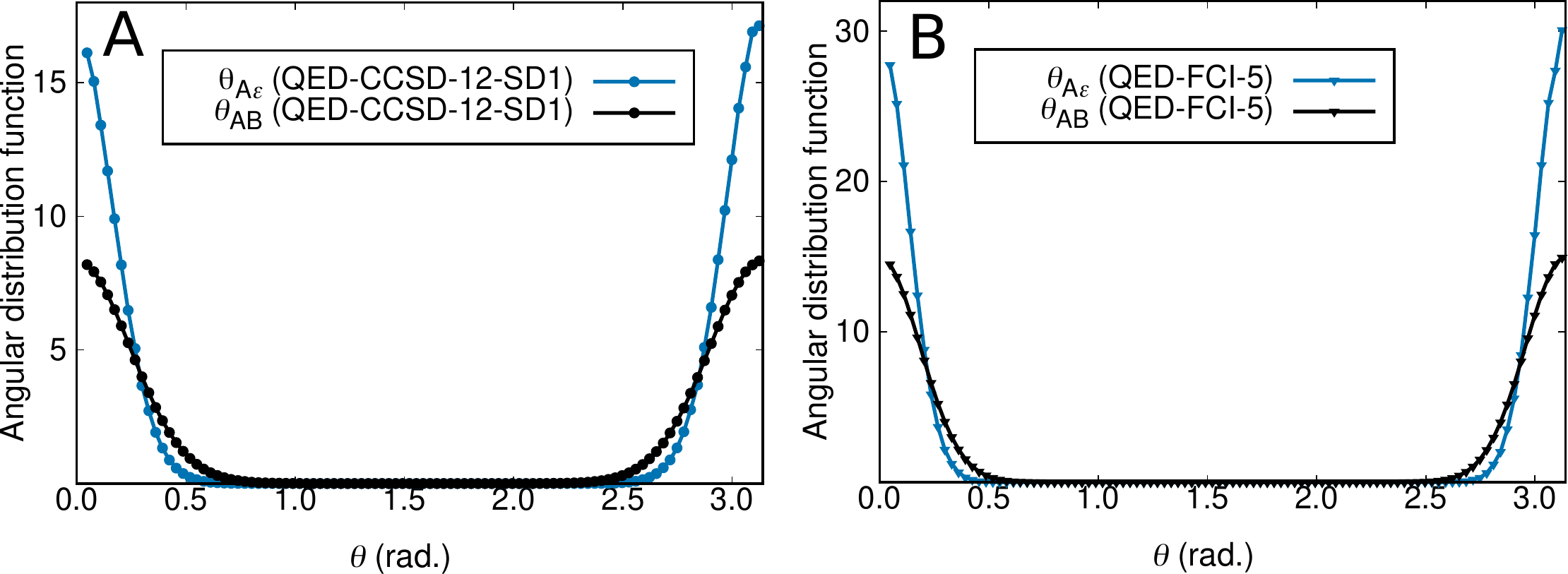}
    \caption{Angular distribution functions of molecular bond axis of molecule $A$ to the molecular bond axis of molecule $B$ ($\theta_{\rm AB}$) and angular distribution functions of molecular bond axis of molecule $A$ to the cavity polarization vector ($\theta_{\rm A\epsilon}$) for $1,000$ H$_2$ molecules of a classical MD trajectory with the NN potentials obtained from training the ML model on (A) QED-CCSD-12-SD1 and (B) QED-FCI-5 data sets.}
    \label{fig:CCSD-2VsFCI}
\end{figure*}

Here we compare our results of classical MD simulations using the ML potentials obtained from QED-FCI-5 and QED-CCSD-12-SD1 calculations. As summarized in~\ref{fig:CCSD-2VsFCI}, we see that classical MD with ML potentials that are obtained from the two different levels of $\it ab~initio$ calculations qualitatively match each other. However, the intensities in the angular distribution functions of $\theta_{\rm A\epsilon}$ and $\theta_{\rm AB}$ for the two cases are different. These differences are due to the quantitative differences in predicting the interaction energies using these two methods (see~\ref{fig:pes-qed-methods}).   

\subsection{$\lambda$ Dependent Molecular Alignment}

Two different $\lambda$ values were considered in our study. In the main text, we focused our discussion on the results with $\lambda = 0.1$~a.u. In this section, we study the properties of a system with $\lambda = 0.02$~a.u. and compare these results with the results obtained using $\lambda = 0.1$~a.u. 

In order to train a model with $\lambda = 0.02$~a.u. important NN parameters for $c_0$ and $c_3$ were transferred and scaled from our training model with $\lambda  = 0.1$~a.u. together with the perturbation theory analysis. The accuracy of the model has been tested by plotting the energies obtained from the NNPs against the {\it ab initio} energies. A linearity plot is obtained as shown in~\ref{fig:cavity_lambda02_pes}A. Additionally, scanned potential energy curves of several selected pair configurations are in good agreement with {\it ab initio} potential energy curves. Some of these plots are shown in~\ref{fig:cavity_lambda02_pes}B. The accuracy of our ML model is further justified with in~\ref{fig:cavity_lambda02_pes}C, where we show that our ML model correctly predicts the long range interaction energy with different directions of the cavity polarization vector. 

\begin{figure*}
    \centering
    \includegraphics[width=17.0cm]{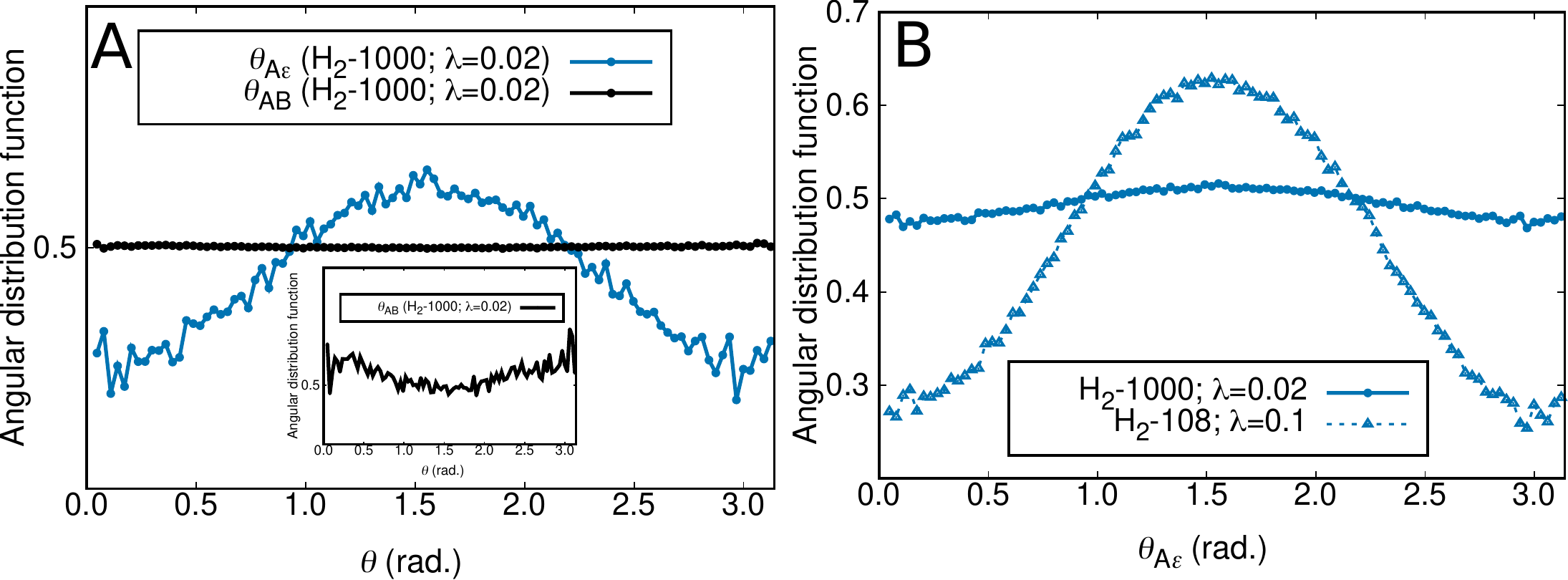}
    \caption{Angular distribution functions of molecular bond axis of molecule $A$ to the molecular bond axis of molecule $B$ ($\theta_{\rm AB}$) and angular distribution functions of molecular bond axis of molecule $A$ to the cavity polarization vector ($\theta_{\rm A\epsilon}$) for $1,000$ H$_2$ molecules of a classical MD trajectory with the NNPs obtained from the training ML model on (A) QED-CCSD-12-SD1 and $\lambda = 0.02$~a.u. coupling constant are shown. A zoom-in figure of $\theta_{\rm A\epsilon}$ is shown in the inset. (B) Angular distribution functions of molecular bond axis of molecule $A$ to the cavity polarization vector ($\theta_{\rm A\epsilon}$) of $108$ molecules with $\lambda = 0.1$~a.u. (dashed line) and $1,000$ molecules with $\lambda = 0.02$~a.u. (solid line) are shown.}
    \label{fig:lambda_dependency}
\end{figure*}

\begin{figure*}
    \centering
    \includegraphics[width=17.0cm]{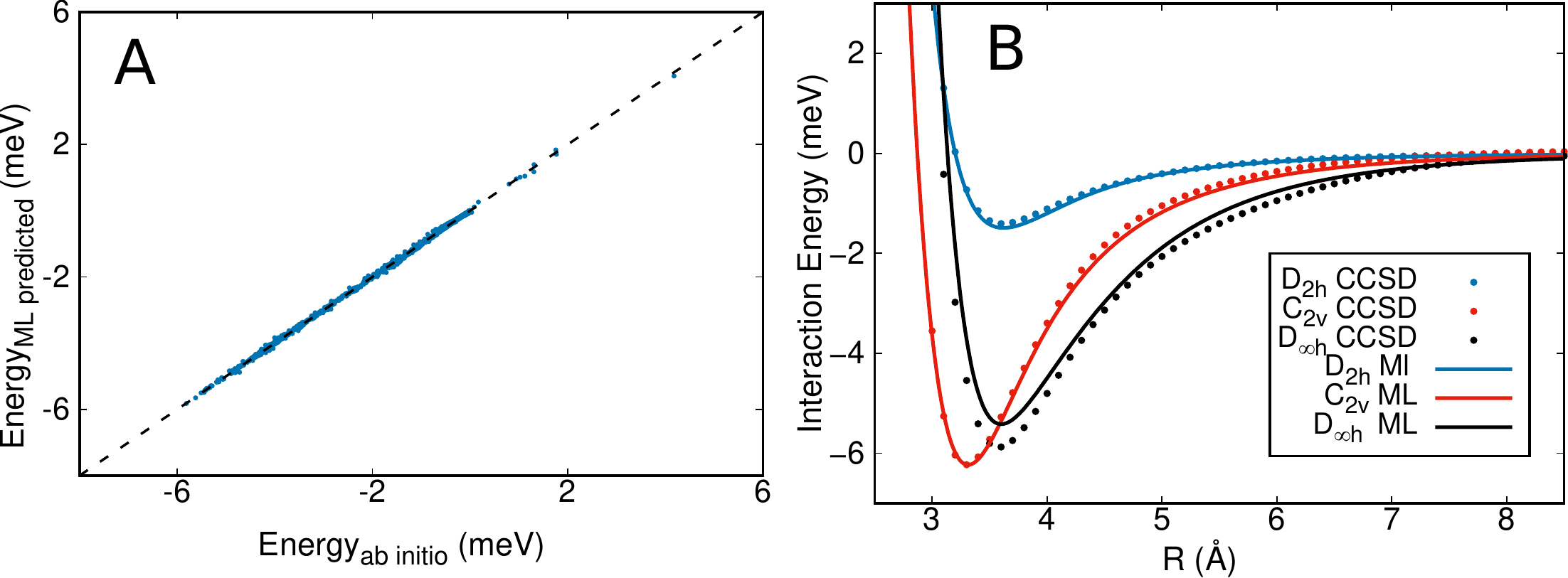}
    \caption{(A) Pairwise interaction energies obtained from {\it ab initio}, CCSD calculation (without cavity) and ML predicted energies are plotted. (B) Scanned potential energy curve for D$_{2\rm h}$, C$_{2\rm v}$ and D$_{\infty h}$ configuration of a pair of molecules using NNPs and from {\it ab initio} calculation are shown. }
    \label{fig:nocavity_pes}
\end{figure*}

\begin{figure*}
    \centering
    \includegraphics[width=17.0cm]{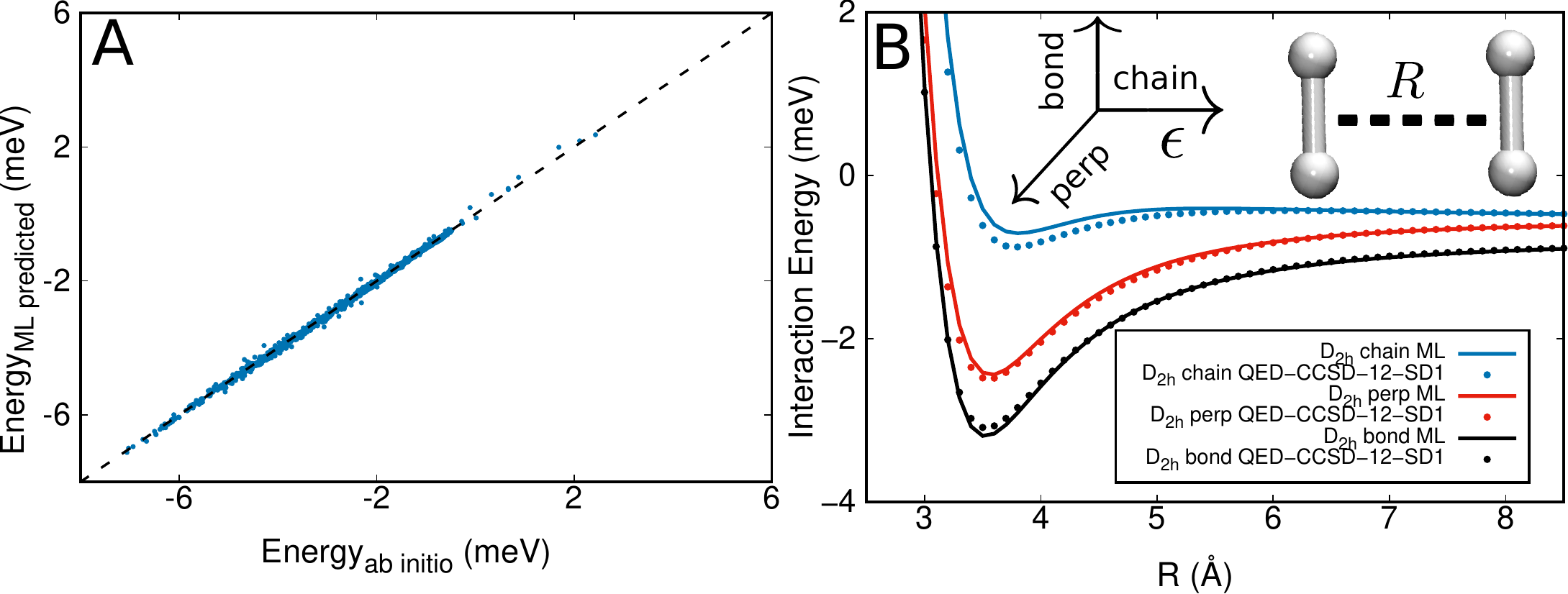}
    \caption{(A) Pairwise interaction energies obtained from {\it ab initio}, QED-CCSD-12-SD1 calculation (with cavity) and ML predicted energies are plotted. (B) Scanned potential energy curve for D$_{2\rm h}$ configuration with three different direction of cavity polarization using NNPs and from {\it ab initio} calculation are shown.}
    \label{fig:cavity_pes}
\end{figure*}

\begin{figure*}
    \centering
    \includegraphics[width=17.0cm]{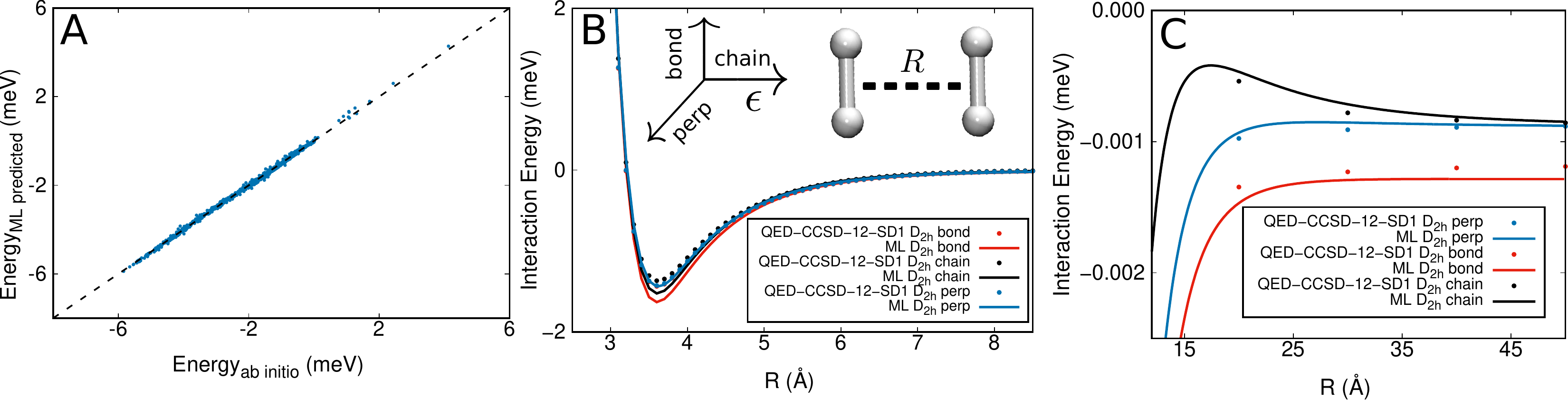}
    \caption{(A) Pairwise interaction energies obtained from {\it ab initio}, QED-CCSD-12-SD1 calculation and ML predicted energies with $\lambda = 0.02$~a.u. are plotted. (B) Scanned potential energy curves for D$_{2\rm h}$ configuration of a pair of molecules using NNPs and from {\it ab initio} calculation are shown. Distance ($R$) between molecule $A$ and molecule $B$ over which potential energy is scanned is shown in the inset of the figure. (C) Scanned potential energy curves for D$_{\rm 2h}$ configuration at the long range are shown. ML model can accurately distinguish different configurations at long distance.}
    \label{fig:cavity_lambda02_pes}
\end{figure*}

\begin{figure}
    \centering
    \includegraphics[angle=-90,width=9.0cm]{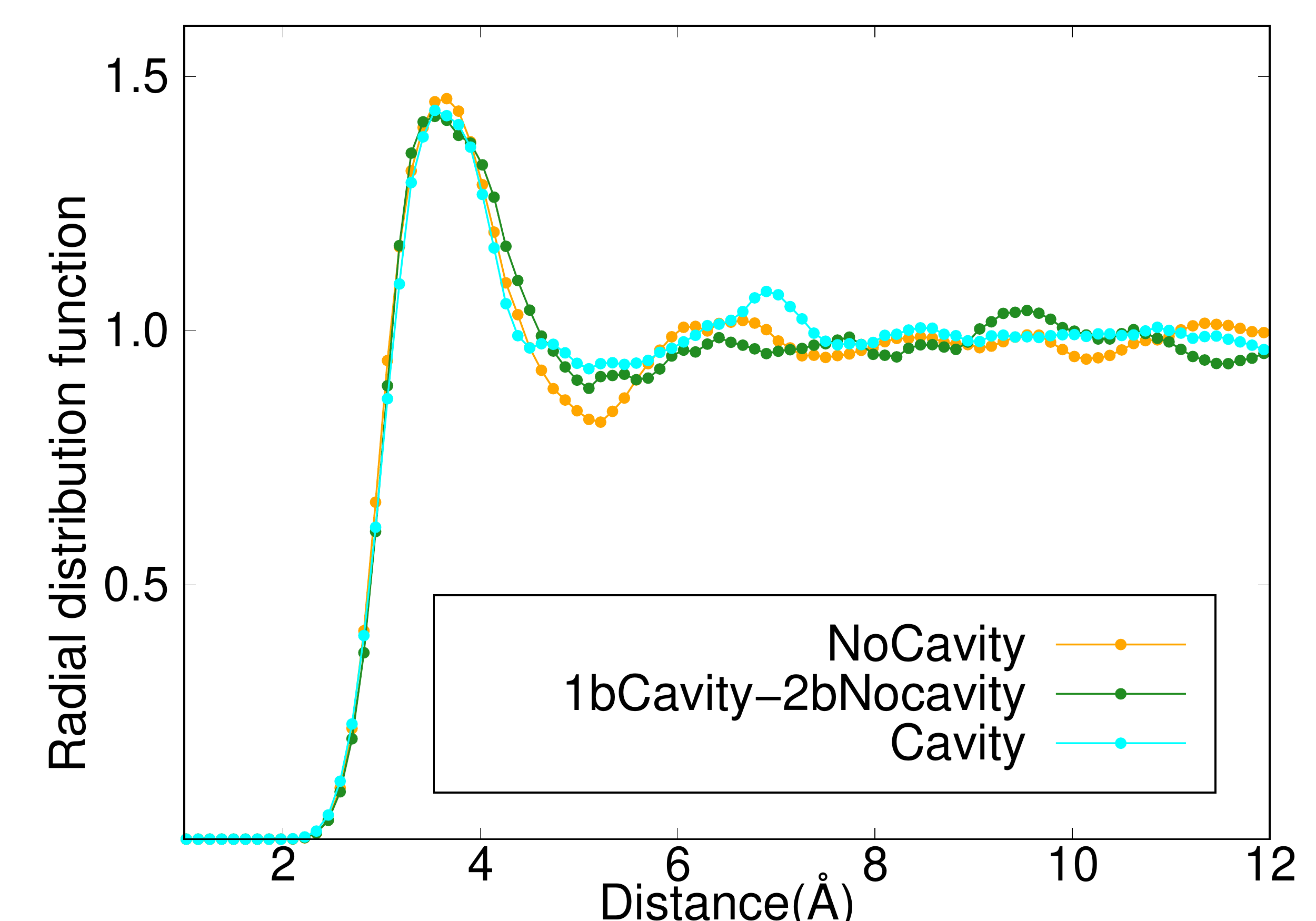}
    \caption{Radial distribution function generated using PIMD trajectory with 1000 H$_2$ molecules using pair potential obtained through a ML training on {\it ab initio} calculation with QED-CCSD-12-SD1 and $\lambda = 0.1$~a.u.}
    \label{fig:RDF}
\end{figure}

A significant difference in the angular distribution functions of $\theta_{\rm A\epsilon}$ is observed when the results of two different $\lambda$ values are compared for $1,000$ H$_2$ molecules. The distribution function of $\theta_{\rm A\epsilon}$ for $1,000$ H$_2$ molecules with $\lambda = 0.02$~a.u. (\ref{fig:lambda_dependency}A) shows molecular alignment perpendicular to the cavity polarization ($\theta_{\rm A\epsilon} = \frac{\pi}{2}$). On the other hand, we observe in~\ref{fig:classicalVsPIMD}A that the angular distribution function of $\theta_{\rm A\epsilon}$ is maximized in the direction of cavity polarization vector ($\theta_{\rm A\epsilon} = 0, \pi$) when $\lambda = 0.1$~a.u. This can be explained from our perturbation theory analysis where we showed that the cavity-modifications to the single molecule energies scale with $\lambda^2$ and the extremely long range pairwise interaction scales with $\lambda^4$. Thus, the importance of the pairwise interaction decreases much faster than the single molecule energy contribution as $\lambda$ decreases. In this particular example of $1,000$ H$_2$ molecules with $\lambda = 0.02$~a.u., the single molecule energy dominates whereas, with $\lambda = 0.1$~a.u., the pairwise interaction energy dominates. $\theta_{\rm AB}$ qualitatively follow the same trend as we observed for $1,000$ H$_2$ molecules with $\lambda = 0.1$~a.u.; however, the intensity of the peak is reduced which suggests a weaker synchronization of molecular orientations. This is shown in the inset of~\ref{fig:lambda_dependency}A.

From the above discussion, we understand that the energy contributions from a single molecule can be altered by (1) changing the number of molecules with a fixed $\lambda$, and (2) changing the value of $\lambda$ for a fix number of molecules. We ran  simulations considering these two possibilities. For the first possibility, we reduced the number of molecules from $1,000$ to $108$ while keeping $\lambda$ equal to $0.1$~a.u., and we compute the angular distribution function for $\theta_{\rm A\epsilon}$. We find that in the $108$ molecule simulation the preferential alignment of the molecules is perpendicular to the cavity polarization vector, which is opposite to the alignment of $1,000$ molecules with $\lambda = 0.1$~a.u. (aligned parallel to the cavity polarization vector). These results are shown in~\ref{fig:classicalVsPIMD}A and~\ref{fig:lambda_dependency}B. For the second possibility, we simulate $1,000$ molecules with a reduced value of $\lambda = 0.02$~a.u. The angular distribution function of $\theta_{\rm A\epsilon}$ in this simulation is qualitatively similar to the results obtained in the first possibility with the molecular alignment perpendicular to the cavity polarization vector (see~\ref{fig:classicalVsPIMD}A and~\ref{fig:lambda_dependency}B). All of our numerical simulation results reported in this section further confirm the conceptual validity of our perturbation theory analysis.

\begin{figure*}[htbp]
    \centering{}\includegraphics[width=\textwidth]{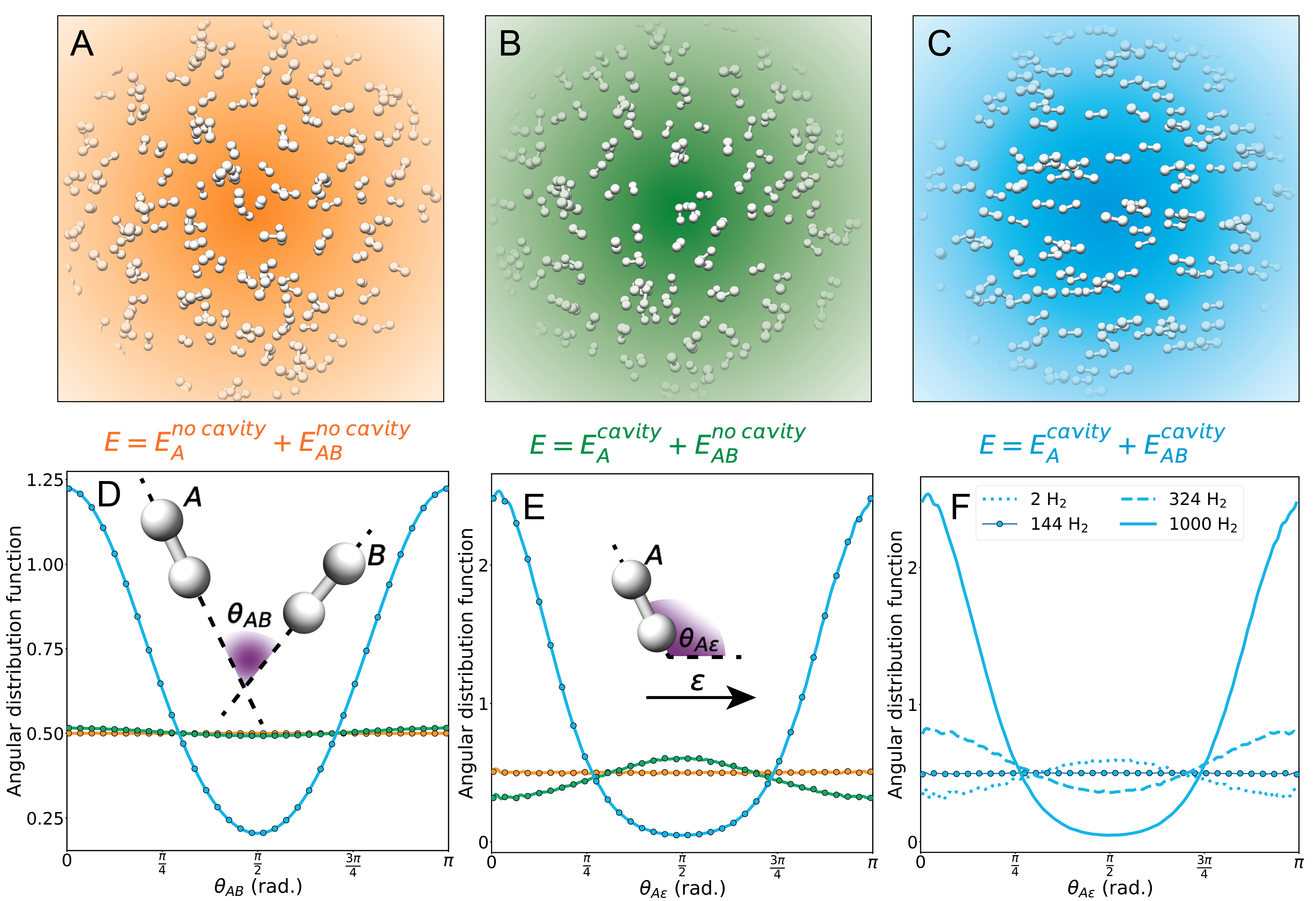} \caption{\label{fig:fig-2-MD}(A-C) Snapshots taken at thermal equilibrium from the path integral molecular dynamic (PIMD) simulations of 1000 H$_2$ molecules in the case of (A) no cavity (orange), (B) cavity-modified one-body term but no cavity two-body term (green), and (C) cavity-modified one-body and two-body terms (blue). For these three cases, the (D) molecular bond axis of molecule $A$ to molecular bond axis of molecule $B$ ($\theta_{AB}$) angular probability distribution function, $P\left(\theta_{AB}\right)$ and (E) molecular bond axis to cavity polarization vector ($\theta_{A\epsilon}$), angular probability distribution function, $P\left(\theta_{A\epsilon}\right)$, are shown. (F) molecular bond axis to cavity polarization vector ($\theta_{A\epsilon}$), angular probability distribution function, $P\left(\theta_{A\epsilon}\right)$, are shown for four different simulations containing different numbers of H$_2$ molecules. All PIMD simulations shown in this figure were performed using neural networks trained with CCSD (no cavity) or QED-CCSD-12-SD1 with $\lambda=0.1$ a.u. (cavity) calculated energies.}
\end{figure*}

\clearpage

\bibliography{cavity} 

\begin{thebibliography}{49}%
\makeatletter
\providecommand \@ifxundefined [1]{%
 \@ifx{#1\undefined}
}%
\providecommand \@ifnum [1]{%
 \ifnum #1\expandafter \@firstoftwo
 \else \expandafter \@secondoftwo
 \fi
}%
\providecommand \@ifx [1]{%
 \ifx #1\expandafter \@firstoftwo
 \else \expandafter \@secondoftwo
 \fi
}%
\providecommand \natexlab [1]{#1}%
\providecommand \enquote  [1]{``#1''}%
\providecommand \bibnamefont  [1]{#1}%
\providecommand \bibfnamefont [1]{#1}%
\providecommand \citenamefont [1]{#1}%
\providecommand \href@noop [0]{\@secondoftwo}%
\providecommand \href [0]{\begingroup \@sanitize@url \@href}%
\providecommand \@href[1]{\@@startlink{#1}\@@href}%
\providecommand \@@href[1]{\endgroup#1\@@endlink}%
\providecommand \@sanitize@url [0]{\catcode `\\12\catcode `\$12\catcode
  `\&12\catcode `\#12\catcode `\^12\catcode `\_12\catcode `\%12\relax}%
\providecommand \@@startlink[1]{}%
\providecommand \@@endlink[0]{}%
\providecommand \url  [0]{\begingroup\@sanitize@url \@url }%
\providecommand \@url [1]{\endgroup\@href {#1}{\urlprefix }}%
\providecommand \urlprefix  [0]{URL }%
\providecommand \Eprint [0]{\href }%
\providecommand \doibase [0]{https://doi.org/}%
\providecommand \selectlanguage [0]{\@gobble}%
\providecommand \bibinfo  [0]{\@secondoftwo}%
\providecommand \bibfield  [0]{\@secondoftwo}%
\providecommand \translation [1]{[#1]}%
\providecommand \BibitemOpen [0]{}%
\providecommand \bibitemStop [0]{}%
\providecommand \bibitemNoStop [0]{.\EOS\space}%
\providecommand \EOS [0]{\spacefactor3000\relax}%
\providecommand \BibitemShut  [1]{\csname bibitem#1\endcsname}%
\let\auto@bib@innerbib\@empty
\bibitem [{\citenamefont {Hobza}\ and\ \citenamefont
  {Šponer}(2002)}]{Hobza2002}%
  \BibitemOpen
  \bibfield  {author} {\bibinfo {author} {\bibfnamefont {P.}~\bibnamefont
  {Hobza}}\ and\ \bibinfo {author} {\bibfnamefont {J.}~\bibnamefont
  {Šponer}},\ }\href {https://doi.org/10.1021/ja026759n} {\bibfield  {journal}
  {\bibinfo  {journal} {J. Am. Chem. Soc.}\ }\textbf {\bibinfo {volume}
  {124}},\ \bibinfo {pages} {11802} (\bibinfo {year} {2002})}\BibitemShut
  {NoStop}%
\bibitem [{\citenamefont {Novoselov}\ \emph {et~al.}(2016)\citenamefont
  {Novoselov}, \citenamefont {Mishchenko}, \citenamefont {Carvalho},\ and\
  \citenamefont {Neto}}]{Novoselov2016}%
  \BibitemOpen
  \bibfield  {author} {\bibinfo {author} {\bibfnamefont {K.~S.}\ \bibnamefont
  {Novoselov}}, \bibinfo {author} {\bibfnamefont {A.}~\bibnamefont
  {Mishchenko}}, \bibinfo {author} {\bibfnamefont {A.}~\bibnamefont
  {Carvalho}},\ and\ \bibinfo {author} {\bibfnamefont {A.~H.~C.}\ \bibnamefont
  {Neto}},\ }\href {https://doi.org/10.1126/science.aac9439} {\bibfield
  {journal} {\bibinfo  {journal} {Science}\ }\textbf {\bibinfo {volume}
  {353}},\ \bibinfo {pages} {aac9439} (\bibinfo {year} {2016})}\BibitemShut
  {NoStop}%
\bibitem [{\citenamefont {Sternbach}\ \emph {et~al.}(2021)\citenamefont
  {Sternbach}, \citenamefont {Chae}, \citenamefont {Latini}, \citenamefont
  {Rikhter}, \citenamefont {Shao}, \citenamefont {Li}, \citenamefont {Rhodes},
  \citenamefont {Kim}, \citenamefont {Schuck}, \citenamefont {Xu},
  \citenamefont {Zhu}, \citenamefont {Averitt}, \citenamefont {Hone},
  \citenamefont {Fogler}, \citenamefont {Rubio},\ and\ \citenamefont
  {Basov}}]{Sternbach2021}%
  \BibitemOpen
  \bibfield  {author} {\bibinfo {author} {\bibfnamefont {A.~J.}\ \bibnamefont
  {Sternbach}}, \bibinfo {author} {\bibfnamefont {S.~H.}\ \bibnamefont {Chae}},
  \bibinfo {author} {\bibfnamefont {S.}~\bibnamefont {Latini}}, \bibinfo
  {author} {\bibfnamefont {A.~A.}\ \bibnamefont {Rikhter}}, \bibinfo {author}
  {\bibfnamefont {Y.}~\bibnamefont {Shao}}, \bibinfo {author} {\bibfnamefont
  {B.}~\bibnamefont {Li}}, \bibinfo {author} {\bibfnamefont {D.}~\bibnamefont
  {Rhodes}}, \bibinfo {author} {\bibfnamefont {B.}~\bibnamefont {Kim}},
  \bibinfo {author} {\bibfnamefont {P.~J.}\ \bibnamefont {Schuck}}, \bibinfo
  {author} {\bibfnamefont {X.}~\bibnamefont {Xu}}, \bibinfo {author}
  {\bibfnamefont {X.~Y.}\ \bibnamefont {Zhu}}, \bibinfo {author} {\bibfnamefont
  {R.~D.}\ \bibnamefont {Averitt}}, \bibinfo {author} {\bibfnamefont
  {J.}~\bibnamefont {Hone}}, \bibinfo {author} {\bibfnamefont {M.~M.}\
  \bibnamefont {Fogler}}, \bibinfo {author} {\bibfnamefont {A.}~\bibnamefont
  {Rubio}},\ and\ \bibinfo {author} {\bibfnamefont {D.~N.}\ \bibnamefont
  {Basov}},\ }\href {https://doi.org/10.1126/science.abe9163} {\bibfield
  {journal} {\bibinfo  {journal} {Science}\ }\textbf {\bibinfo {volume}
  {371}},\ \bibinfo {pages} {617} (\bibinfo {year} {2021})}\BibitemShut
  {NoStop}%
\bibitem [{\citenamefont {Maitland}\ \emph {et~al.}(1981)\citenamefont
  {Maitland}, \citenamefont {Maitland}, \citenamefont {Rigby}, \citenamefont
  {Smith},\ and\ \citenamefont {Wakeham}}]{Maitland1981}%
  \BibitemOpen
  \bibfield  {author} {\bibinfo {author} {\bibfnamefont {G.~C.}\ \bibnamefont
  {Maitland}}, \bibinfo {author} {\bibfnamefont {G.~D.}\ \bibnamefont
  {Maitland}}, \bibinfo {author} {\bibfnamefont {M.}~\bibnamefont {Rigby}},
  \bibinfo {author} {\bibfnamefont {E.~B.}\ \bibnamefont {Smith}},\ and\
  \bibinfo {author} {\bibfnamefont {W.~A.}\ \bibnamefont {Wakeham}},\ }\href
  {https://books.google.com/books?id=qhHwAAAAMAAJ} {\emph {\bibinfo {title}
  {{Intermolecular Forces: Their Origin and Determination}}}}\ (\bibinfo
  {publisher} {Oxford University Press, USA},\ \bibinfo {year}
  {1981})\BibitemShut {NoStop}%
\bibitem [{\citenamefont {Stone}(2013)}]{Stone2013}%
  \BibitemOpen
  \bibfield  {author} {\bibinfo {author} {\bibfnamefont {A.}~\bibnamefont
  {Stone}},\ }\href {https://doi.org/10.1093/acprof:oso/9780199672394.001.0001}
  {\emph {\bibinfo {title} {{The Theory of Intermolecular Forces}}}},\ \bibinfo
  {edition} {2nd}\ ed.\ (\bibinfo  {publisher} {Oxford University Press},\
  \bibinfo {address} {Oxford},\ \bibinfo {year} {2013})\ p.\ \bibinfo {pages}
  {352}\BibitemShut {NoStop}%
\bibitem [{\citenamefont {London}(1937)}]{London1937}%
  \BibitemOpen
  \bibfield  {author} {\bibinfo {author} {\bibfnamefont {F.}~\bibnamefont
  {London}},\ }\href {https://doi.org/10.1039/TF937330008B} {\bibfield
  {journal} {\bibinfo  {journal} {Trans. Faraday Soc.}\ }\textbf {\bibinfo
  {volume} {33}},\ \bibinfo {pages} {8b} (\bibinfo {year} {1937})}\BibitemShut
  {NoStop}%
\bibitem [{\citenamefont {Halgren}(1992)}]{Halgren1992}%
  \BibitemOpen
  \bibfield  {author} {\bibinfo {author} {\bibfnamefont {T.~A.}\ \bibnamefont
  {Halgren}},\ }\href {https://doi.org/10.1021/ja00046a032} {\bibfield
  {journal} {\bibinfo  {journal} {J. Am. Chem. Soc.}\ }\textbf {\bibinfo
  {volume} {114}},\ \bibinfo {pages} {7827} (\bibinfo {year}
  {1992})}\BibitemShut {NoStop}%
\bibitem [{\citenamefont {Grimme}\ \emph {et~al.}(2010)\citenamefont {Grimme},
  \citenamefont {Antony}, \citenamefont {Ehrlich},\ and\ \citenamefont
  {Krieg}}]{Grimme2010}%
  \BibitemOpen
  \bibfield  {author} {\bibinfo {author} {\bibfnamefont {S.}~\bibnamefont
  {Grimme}}, \bibinfo {author} {\bibfnamefont {J.}~\bibnamefont {Antony}},
  \bibinfo {author} {\bibfnamefont {S.}~\bibnamefont {Ehrlich}},\ and\ \bibinfo
  {author} {\bibfnamefont {H.}~\bibnamefont {Krieg}},\ }\href
  {https://doi.org/10.1063/1.3382344} {\bibfield  {journal} {\bibinfo
  {journal} {J. Chem. Phys.}\ }\textbf {\bibinfo {volume} {132}},\ \bibinfo
  {pages} {154104} (\bibinfo {year} {2010})}\BibitemShut {NoStop}%
\bibitem [{\citenamefont {Thirunamachandran}(1980)}]{Thirunamachandran1980}%
  \BibitemOpen
  \bibfield  {author} {\bibinfo {author} {\bibfnamefont {T.}~\bibnamefont
  {Thirunamachandran}},\ }\href {https://doi.org/10.1080/00268978000101561}
  {\bibfield  {journal} {\bibinfo  {journal} {Mol. Phys.}\ }\textbf {\bibinfo
  {volume} {40}},\ \bibinfo {pages} {393} (\bibinfo {year} {1980})}\BibitemShut
  {NoStop}%
\bibitem [{\citenamefont {Milonni}\ and\ \citenamefont
  {Smith}(1996)}]{Milonni1996}%
  \BibitemOpen
  \bibfield  {author} {\bibinfo {author} {\bibfnamefont {P.~W.}\ \bibnamefont
  {Milonni}}\ and\ \bibinfo {author} {\bibfnamefont {A.}~\bibnamefont
  {Smith}},\ }\href {https://doi.org/10.1103/PhysRevA.53.3484} {\bibfield
  {journal} {\bibinfo  {journal} {Phys. Rev. A}\ }\textbf {\bibinfo {volume}
  {53}},\ \bibinfo {pages} {3484} (\bibinfo {year} {1996})}\BibitemShut
  {NoStop}%
\bibitem [{\citenamefont {Sherkunov}(2009)}]{Sherkunov2009}%
  \BibitemOpen
  \bibfield  {author} {\bibinfo {author} {\bibfnamefont {Y.}~\bibnamefont
  {Sherkunov}},\ }\href {https://doi.org/10.1088/1742-6596/161/1/012041}
  {\bibfield  {journal} {\bibinfo  {journal} {J. Phys. Conf. Ser.}\ }\textbf
  {\bibinfo {volume} {161}},\ \bibinfo {pages} {012041} (\bibinfo {year}
  {2009})}\BibitemShut {NoStop}%
\bibitem [{\citenamefont {Fiscelli}\ \emph {et~al.}(2020)\citenamefont
  {Fiscelli}, \citenamefont {Rizzuto},\ and\ \citenamefont
  {Passante}}]{Fiscelli2020}%
  \BibitemOpen
  \bibfield  {author} {\bibinfo {author} {\bibfnamefont {G.}~\bibnamefont
  {Fiscelli}}, \bibinfo {author} {\bibfnamefont {L.}~\bibnamefont {Rizzuto}},\
  and\ \bibinfo {author} {\bibfnamefont {R.}~\bibnamefont {Passante}},\ }\href
  {https://doi.org/10.1103/PhysRevLett.124.013604} {\bibfield  {journal}
  {\bibinfo  {journal} {Phys. Rev. Lett.}\ }\textbf {\bibinfo {volume} {124}},\
  \bibinfo {pages} {013604} (\bibinfo {year} {2020})}\BibitemShut {NoStop}%
\bibitem [{\citenamefont {Haugland}\ \emph {et~al.}(2021)\citenamefont
  {Haugland}, \citenamefont {Sch{\"{a}}fer}, \citenamefont {Ronca},
  \citenamefont {Rubio},\ and\ \citenamefont {Koch}}]{Haugland2021}%
  \BibitemOpen
  \bibfield  {author} {\bibinfo {author} {\bibfnamefont {T.~S.}\ \bibnamefont
  {Haugland}}, \bibinfo {author} {\bibfnamefont {C.}~\bibnamefont
  {Sch{\"{a}}fer}}, \bibinfo {author} {\bibfnamefont {E.}~\bibnamefont
  {Ronca}}, \bibinfo {author} {\bibfnamefont {A.}~\bibnamefont {Rubio}},\ and\
  \bibinfo {author} {\bibfnamefont {H.}~\bibnamefont {Koch}},\ }\href
  {https://doi.org/10.1063/5.0039256} {\bibfield  {journal} {\bibinfo
  {journal} {J. Chem. Phys.}\ }\textbf {\bibinfo {volume} {154}},\ \bibinfo
  {pages} {094113} (\bibinfo {year} {2021})}\BibitemShut {NoStop}%
\bibitem [{\citenamefont {Ribeiro}\ \emph {et~al.}(2018)\citenamefont
  {Ribeiro}, \citenamefont {Mart{\'{i}}nez-Mart{\'{i}}nez}, \citenamefont {Du},
  \citenamefont {Campos-Gonzalez-Angulo},\ and\ \citenamefont
  {Yuen-Zhou}}]{RibeiroChemSci2018}%
  \BibitemOpen
  \bibfield  {author} {\bibinfo {author} {\bibfnamefont {R.~F.}\ \bibnamefont
  {Ribeiro}}, \bibinfo {author} {\bibfnamefont {L.~A.}\ \bibnamefont
  {Mart{\'{i}}nez-Mart{\'{i}}nez}}, \bibinfo {author} {\bibfnamefont
  {M.}~\bibnamefont {Du}}, \bibinfo {author} {\bibfnamefont {J.}~\bibnamefont
  {Campos-Gonzalez-Angulo}},\ and\ \bibinfo {author} {\bibfnamefont
  {J.}~\bibnamefont {Yuen-Zhou}},\ }\href {https://doi.org/10.1039/C8SC01043A}
  {\bibfield  {journal} {\bibinfo  {journal} {Chem. Sci.}\ }\textbf {\bibinfo
  {volume} {9}},\ \bibinfo {pages} {6325} (\bibinfo {year} {2018})}\BibitemShut
  {NoStop}%
\bibitem [{\citenamefont {Rivera}\ \emph {et~al.}(2019)\citenamefont {Rivera},
  \citenamefont {Flick},\ and\ \citenamefont {Narang}}]{Rivera2019}%
  \BibitemOpen
  \bibfield  {author} {\bibinfo {author} {\bibfnamefont {N.}~\bibnamefont
  {Rivera}}, \bibinfo {author} {\bibfnamefont {J.}~\bibnamefont {Flick}},\ and\
  \bibinfo {author} {\bibfnamefont {P.}~\bibnamefont {Narang}},\ }\href
  {https://doi.org/10.1103/PhysRevLett.122.193603} {\bibfield  {journal}
  {\bibinfo  {journal} {Phys. Rev. Lett.}\ }\textbf {\bibinfo {volume} {122}},\
  \bibinfo {pages} {193603} (\bibinfo {year} {2019})}\BibitemShut {NoStop}%
\bibitem [{\citenamefont {Thomas}\ \emph {et~al.}(2019)\citenamefont {Thomas},
  \citenamefont {Lethuillier-Karl}, \citenamefont {Nagarajan}, \citenamefont
  {Vergauwe}, \citenamefont {George}, \citenamefont {Chervy}, \citenamefont
  {Shalabney}, \citenamefont {Devaux}, \citenamefont {Genet}, \citenamefont
  {Moran},\ and\ \citenamefont {Ebbesen}}]{ThomasScience2019}%
  \BibitemOpen
  \bibfield  {author} {\bibinfo {author} {\bibfnamefont {A.}~\bibnamefont
  {Thomas}}, \bibinfo {author} {\bibfnamefont {L.}~\bibnamefont
  {Lethuillier-Karl}}, \bibinfo {author} {\bibfnamefont {K.}~\bibnamefont
  {Nagarajan}}, \bibinfo {author} {\bibfnamefont {R.~M.~A.}\ \bibnamefont
  {Vergauwe}}, \bibinfo {author} {\bibfnamefont {J.}~\bibnamefont {George}},
  \bibinfo {author} {\bibfnamefont {T.}~\bibnamefont {Chervy}}, \bibinfo
  {author} {\bibfnamefont {A.}~\bibnamefont {Shalabney}}, \bibinfo {author}
  {\bibfnamefont {E.}~\bibnamefont {Devaux}}, \bibinfo {author} {\bibfnamefont
  {C.}~\bibnamefont {Genet}}, \bibinfo {author} {\bibfnamefont
  {J.}~\bibnamefont {Moran}},\ and\ \bibinfo {author} {\bibfnamefont {T.~W.}\
  \bibnamefont {Ebbesen}},\ }\href {https://doi.org/10.1126/science.aau7742}
  {\bibfield  {journal} {\bibinfo  {journal} {Science}\ }\textbf {\bibinfo
  {volume} {363}},\ \bibinfo {pages} {615} (\bibinfo {year}
  {2019})}\BibitemShut {NoStop}%
\bibitem [{\citenamefont {Li}\ \emph {et~al.}(2020)\citenamefont {Li},
  \citenamefont {Subotnik},\ and\ \citenamefont {Nitzan}}]{Li2020a}%
  \BibitemOpen
  \bibfield  {author} {\bibinfo {author} {\bibfnamefont {T.~E.}\ \bibnamefont
  {Li}}, \bibinfo {author} {\bibfnamefont {J.~E.}\ \bibnamefont {Subotnik}},\
  and\ \bibinfo {author} {\bibfnamefont {A.}~\bibnamefont {Nitzan}},\ }\href
  {https://doi.org/10.1073/pnas.2009272117} {\bibfield  {journal} {\bibinfo
  {journal} {Proc. Natl. Acad. Sci. U. S. A.}\ }\textbf {\bibinfo {volume}
  {117}},\ \bibinfo {pages} {18324} (\bibinfo {year} {2020})}\BibitemShut
  {NoStop}%
\bibitem [{\citenamefont {Garcia-Vidal}\ \emph {et~al.}(2021)\citenamefont
  {Garcia-Vidal}, \citenamefont {Ciuti},\ and\ \citenamefont
  {Ebbesen}}]{Garcia-Vidal2021}%
  \BibitemOpen
  \bibfield  {author} {\bibinfo {author} {\bibfnamefont {F.~J.}\ \bibnamefont
  {Garcia-Vidal}}, \bibinfo {author} {\bibfnamefont {C.}~\bibnamefont
  {Ciuti}},\ and\ \bibinfo {author} {\bibfnamefont {T.~W.}\ \bibnamefont
  {Ebbesen}},\ }\href {https://doi.org/10.1126/science.abd0336} {\bibfield
  {journal} {\bibinfo  {journal} {Science}\ }\textbf {\bibinfo {volume}
  {373}},\ \bibinfo {pages} {eabd0336} (\bibinfo {year} {2021})}\BibitemShut
  {NoStop}%
\bibitem [{\citenamefont {Li}\ \emph {et~al.}(2021{\natexlab{a}})\citenamefont
  {Li}, \citenamefont {Nitzan},\ and\ \citenamefont {Subotnik}}]{Li2021}%
  \BibitemOpen
  \bibfield  {author} {\bibinfo {author} {\bibfnamefont {T.~E.}\ \bibnamefont
  {Li}}, \bibinfo {author} {\bibfnamefont {A.}~\bibnamefont {Nitzan}},\ and\
  \bibinfo {author} {\bibfnamefont {J.~E.}\ \bibnamefont {Subotnik}},\ }\href
  {https://doi.org/10.1002/ange.202103920} {\bibfield  {journal} {\bibinfo
  {journal} {Angew. Chemie}\ }\textbf {\bibinfo {volume} {133}},\ \bibinfo
  {pages} {15661} (\bibinfo {year} {2021}{\natexlab{a}})}\BibitemShut {NoStop}%
\bibitem [{\citenamefont {Vahala}(2003)}]{Vahala2003}%
  \BibitemOpen
  \bibfield  {author} {\bibinfo {author} {\bibfnamefont {K.~J.}\ \bibnamefont
  {Vahala}},\ }\href {https://doi.org/10.1038/nature01939} {\bibfield
  {journal} {\bibinfo  {journal} {Nature}\ }\textbf {\bibinfo {volume} {424}},\
  \bibinfo {pages} {839} (\bibinfo {year} {2003})}\BibitemShut {NoStop}%
\bibitem [{\citenamefont {Cortese}\ \emph {et~al.}(2017)\citenamefont
  {Cortese}, \citenamefont {Lagoudakis},\ and\ \citenamefont
  {De~Liberato}}]{deLiberato2017}%
  \BibitemOpen
  \bibfield  {author} {\bibinfo {author} {\bibfnamefont {E.}~\bibnamefont
  {Cortese}}, \bibinfo {author} {\bibfnamefont {P.~G.}\ \bibnamefont
  {Lagoudakis}},\ and\ \bibinfo {author} {\bibfnamefont {S.}~\bibnamefont
  {De~Liberato}},\ }\href {https://doi.org/10.1103/PhysRevLett.119.043604}
  {\bibfield  {journal} {\bibinfo  {journal} {Phys. Rev. Lett.}\ }\textbf
  {\bibinfo {volume} {119}},\ \bibinfo {pages} {043604} (\bibinfo {year}
  {2017})}\BibitemShut {NoStop}%
\bibitem [{\citenamefont {Joseph}\ \emph {et~al.}(2021)\citenamefont {Joseph},
  \citenamefont {Kushida}, \citenamefont {Smarsly}, \citenamefont {Ihiawakrim},
  \citenamefont {Thomas}, \citenamefont {Paravicini-Bagliani}, \citenamefont
  {Nagarajan}, \citenamefont {Vergauwe}, \citenamefont {Devaux}, \citenamefont
  {Ersen}, \citenamefont {Bunz},\ and\ \citenamefont {Ebbesen}}]{Joseph2021}%
  \BibitemOpen
  \bibfield  {author} {\bibinfo {author} {\bibfnamefont {K.}~\bibnamefont
  {Joseph}}, \bibinfo {author} {\bibfnamefont {S.}~\bibnamefont {Kushida}},
  \bibinfo {author} {\bibfnamefont {E.}~\bibnamefont {Smarsly}}, \bibinfo
  {author} {\bibfnamefont {D.}~\bibnamefont {Ihiawakrim}}, \bibinfo {author}
  {\bibfnamefont {A.}~\bibnamefont {Thomas}}, \bibinfo {author} {\bibfnamefont
  {G.~L.}\ \bibnamefont {Paravicini-Bagliani}}, \bibinfo {author}
  {\bibfnamefont {K.}~\bibnamefont {Nagarajan}}, \bibinfo {author}
  {\bibfnamefont {R.}~\bibnamefont {Vergauwe}}, \bibinfo {author}
  {\bibfnamefont {E.}~\bibnamefont {Devaux}}, \bibinfo {author} {\bibfnamefont
  {O.}~\bibnamefont {Ersen}}, \bibinfo {author} {\bibfnamefont {U.~H.~F.}\
  \bibnamefont {Bunz}},\ and\ \bibinfo {author} {\bibfnamefont {T.~W.}\
  \bibnamefont {Ebbesen}},\ }\href {https://doi.org/10.1002/anie.202105840}
  {\bibfield  {journal} {\bibinfo  {journal} {Angew. Chem. Int. Ed.}\ }\textbf
  {\bibinfo {volume} {60}},\ \bibinfo {pages} {19665} (\bibinfo {year}
  {2021})}\BibitemShut {NoStop}%
\bibitem [{\citenamefont {Fukushima}\ \emph {et~al.}(2022)\citenamefont
  {Fukushima}, \citenamefont {Yoshimitsu},\ and\ \citenamefont
  {Murakoshi}}]{Fukushima2022}%
  \BibitemOpen
  \bibfield  {author} {\bibinfo {author} {\bibfnamefont {T.}~\bibnamefont
  {Fukushima}}, \bibinfo {author} {\bibfnamefont {S.}~\bibnamefont
  {Yoshimitsu}},\ and\ \bibinfo {author} {\bibfnamefont {K.}~\bibnamefont
  {Murakoshi}},\ }\href {https://doi.org/10.1021/jacs.2c02991} {\bibfield
  {journal} {\bibinfo  {journal} {J. Am. Chem. Soc.}\ }\textbf {\bibinfo
  {volume} {144}},\ \bibinfo {pages} {12177} (\bibinfo {year}
  {2022})}\BibitemShut {NoStop}%
\bibitem [{\citenamefont {Sandeep}\ \emph {et~al.}(2022)\citenamefont
  {Sandeep}, \citenamefont {Joseph}, \citenamefont {Gautier}, \citenamefont
  {Nagarajan}, \citenamefont {Sujith}, \citenamefont {Thomas},\ and\
  \citenamefont {Ebbesen}}]{Sandeep2022}%
  \BibitemOpen
  \bibfield  {author} {\bibinfo {author} {\bibfnamefont {K.}~\bibnamefont
  {Sandeep}}, \bibinfo {author} {\bibfnamefont {K.}~\bibnamefont {Joseph}},
  \bibinfo {author} {\bibfnamefont {J.}~\bibnamefont {Gautier}}, \bibinfo
  {author} {\bibfnamefont {K.}~\bibnamefont {Nagarajan}}, \bibinfo {author}
  {\bibfnamefont {M.}~\bibnamefont {Sujith}}, \bibinfo {author} {\bibfnamefont
  {K.~G.}\ \bibnamefont {Thomas}},\ and\ \bibinfo {author} {\bibfnamefont
  {T.~W.}\ \bibnamefont {Ebbesen}},\ }\href
  {https://doi.org/10.1021/acs.jpclett.1c03893} {\bibfield  {journal} {\bibinfo
   {journal} {J. Phys. Chem. Lett.}\ }\textbf {\bibinfo {volume} {13}},\
  \bibinfo {pages} {1209} (\bibinfo {year} {2022})}\BibitemShut {NoStop}%
\bibitem [{\citenamefont {Galego}\ \emph {et~al.}(2015)\citenamefont {Galego},
  \citenamefont {Garcia-Vidal},\ and\ \citenamefont
  {Feist}}]{GalegoPhysRevX2015}%
  \BibitemOpen
  \bibfield  {author} {\bibinfo {author} {\bibfnamefont {J.}~\bibnamefont
  {Galego}}, \bibinfo {author} {\bibfnamefont {F.~J.}\ \bibnamefont
  {Garcia-Vidal}},\ and\ \bibinfo {author} {\bibfnamefont {J.}~\bibnamefont
  {Feist}},\ }\href {https://doi.org/10.1103/PhysRevX.5.041022} {\bibfield
  {journal} {\bibinfo  {journal} {Phys. Rev. X}\ }\textbf {\bibinfo {volume}
  {5}},\ \bibinfo {pages} {41022} (\bibinfo {year} {2015})}\BibitemShut
  {NoStop}%
\bibitem [{\citenamefont {Lacombe}\ \emph {et~al.}(2019)\citenamefont
  {Lacombe}, \citenamefont {Hoffmann},\ and\ \citenamefont
  {Maitra}}]{Lacombe2019}%
  \BibitemOpen
  \bibfield  {author} {\bibinfo {author} {\bibfnamefont {L.}~\bibnamefont
  {Lacombe}}, \bibinfo {author} {\bibfnamefont {N.~M.}\ \bibnamefont
  {Hoffmann}},\ and\ \bibinfo {author} {\bibfnamefont {N.~T.}\ \bibnamefont
  {Maitra}},\ }\href {https://doi.org/10.1103/PhysRevLett.123.083201}
  {\bibfield  {journal} {\bibinfo  {journal} {Phys. Rev. Lett.}\ }\textbf
  {\bibinfo {volume} {123}},\ \bibinfo {pages} {083201} (\bibinfo {year}
  {2019})}\BibitemShut {NoStop}%
\bibitem [{\citenamefont {Fregoni}\ \emph {et~al.}(2022)\citenamefont
  {Fregoni}, \citenamefont {Garcia-Vidal},\ and\ \citenamefont
  {Feist}}]{Fregoni2022}%
  \BibitemOpen
  \bibfield  {author} {\bibinfo {author} {\bibfnamefont {J.}~\bibnamefont
  {Fregoni}}, \bibinfo {author} {\bibfnamefont {F.~J.}\ \bibnamefont
  {Garcia-Vidal}},\ and\ \bibinfo {author} {\bibfnamefont {J.}~\bibnamefont
  {Feist}},\ }\href {https://doi.org/10.1021/acsphotonics.1c01749} {\bibfield
  {journal} {\bibinfo  {journal} {ACS Photonics}\ }\textbf {\bibinfo {volume}
  {9}},\ \bibinfo {pages} {1096} (\bibinfo {year} {2022})}\BibitemShut
  {NoStop}%
\bibitem [{\citenamefont {Haugland}\ \emph {et~al.}(2020)\citenamefont
  {Haugland}, \citenamefont {Ronca}, \citenamefont {Kj{\o}nstad}, \citenamefont
  {Rubio},\ and\ \citenamefont {Koch}}]{Haugland2020}%
  \BibitemOpen
  \bibfield  {author} {\bibinfo {author} {\bibfnamefont {T.~S.}\ \bibnamefont
  {Haugland}}, \bibinfo {author} {\bibfnamefont {E.}~\bibnamefont {Ronca}},
  \bibinfo {author} {\bibfnamefont {E.~F.}\ \bibnamefont {Kj{\o}nstad}},
  \bibinfo {author} {\bibfnamefont {A.}~\bibnamefont {Rubio}},\ and\ \bibinfo
  {author} {\bibfnamefont {H.}~\bibnamefont {Koch}},\ }\href
  {https://doi.org/10.1103/PhysRevX.10.041043} {\bibfield  {journal} {\bibinfo
  {journal} {Phys. Rev. X}\ }\textbf {\bibinfo {volume} {10}},\ \bibinfo
  {pages} {041043} (\bibinfo {year} {2020})}\BibitemShut {NoStop}%
\bibitem [{\citenamefont {Zhong}\ \emph {et~al.}(2016)\citenamefont {Zhong},
  \citenamefont {Chervy}, \citenamefont {Wang}, \citenamefont {George},
  \citenamefont {Thomas}, \citenamefont {Hutchison}, \citenamefont {Devaux},
  \citenamefont {Genet},\ and\ \citenamefont {Ebbesen}}]{Zhong2016}%
  \BibitemOpen
  \bibfield  {author} {\bibinfo {author} {\bibfnamefont {X.}~\bibnamefont
  {Zhong}}, \bibinfo {author} {\bibfnamefont {T.}~\bibnamefont {Chervy}},
  \bibinfo {author} {\bibfnamefont {S.}~\bibnamefont {Wang}}, \bibinfo {author}
  {\bibfnamefont {J.}~\bibnamefont {George}}, \bibinfo {author} {\bibfnamefont
  {A.}~\bibnamefont {Thomas}}, \bibinfo {author} {\bibfnamefont {J.~A.}\
  \bibnamefont {Hutchison}}, \bibinfo {author} {\bibfnamefont {E.}~\bibnamefont
  {Devaux}}, \bibinfo {author} {\bibfnamefont {C.}~\bibnamefont {Genet}},\ and\
  \bibinfo {author} {\bibfnamefont {T.~W.}\ \bibnamefont {Ebbesen}},\ }\href
  {https://doi.org/10.1002/anie.201600428} {\bibfield  {journal} {\bibinfo
  {journal} {Angew. Chem. Int. Ed.}\ }\textbf {\bibinfo {volume} {55}},\
  \bibinfo {pages} {6202} (\bibinfo {year} {2016})}\BibitemShut {NoStop}%
\bibitem [{\citenamefont {Du}\ \emph {et~al.}(2018)\citenamefont {Du},
  \citenamefont {Mart{\'{i}}nez-Mart{\'{i}}nez}, \citenamefont {Ribeiro},
  \citenamefont {Hu}, \citenamefont {Menon},\ and\ \citenamefont
  {Yuen-Zhou}}]{DuChemSci2018}%
  \BibitemOpen
  \bibfield  {author} {\bibinfo {author} {\bibfnamefont {M.}~\bibnamefont
  {Du}}, \bibinfo {author} {\bibfnamefont {L.~A.}\ \bibnamefont
  {Mart{\'{i}}nez-Mart{\'{i}}nez}}, \bibinfo {author} {\bibfnamefont {R.~F.}\
  \bibnamefont {Ribeiro}}, \bibinfo {author} {\bibfnamefont {Z.}~\bibnamefont
  {Hu}}, \bibinfo {author} {\bibfnamefont {V.~M.}\ \bibnamefont {Menon}},\ and\
  \bibinfo {author} {\bibfnamefont {J.}~\bibnamefont {Yuen-Zhou}},\ }\href
  {https://doi.org/10.1039/C8SC00171E} {\bibfield  {journal} {\bibinfo
  {journal} {Chem. Sci.}\ }\textbf {\bibinfo {volume} {9}},\ \bibinfo {pages}
  {6659} (\bibinfo {year} {2018})}\BibitemShut {NoStop}%
\bibitem [{\citenamefont {Xiang}\ \emph {et~al.}(2020)\citenamefont {Xiang},
  \citenamefont {Ribeiro}, \citenamefont {Du}, \citenamefont {Chen},
  \citenamefont {Yang}, \citenamefont {Wang}, \citenamefont {Yuen-Zhou},\ and\
  \citenamefont {Xiong}}]{Xiang2020}%
  \BibitemOpen
  \bibfield  {author} {\bibinfo {author} {\bibfnamefont {B.}~\bibnamefont
  {Xiang}}, \bibinfo {author} {\bibfnamefont {R.~F.}\ \bibnamefont {Ribeiro}},
  \bibinfo {author} {\bibfnamefont {M.}~\bibnamefont {Du}}, \bibinfo {author}
  {\bibfnamefont {L.}~\bibnamefont {Chen}}, \bibinfo {author} {\bibfnamefont
  {Z.}~\bibnamefont {Yang}}, \bibinfo {author} {\bibfnamefont {J.}~\bibnamefont
  {Wang}}, \bibinfo {author} {\bibfnamefont {J.}~\bibnamefont {Yuen-Zhou}},\
  and\ \bibinfo {author} {\bibfnamefont {W.}~\bibnamefont {Xiong}},\ }\href
  {https://doi.org/10.1126/science.aba3544} {\bibfield  {journal} {\bibinfo
  {journal} {Science}\ }\textbf {\bibinfo {volume} {368}},\ \bibinfo {pages}
  {665} (\bibinfo {year} {2020})}\BibitemShut {NoStop}%
\bibitem [{\citenamefont {Herrera}\ and\ \citenamefont
  {Spano}(2016)}]{herrera2016}%
  \BibitemOpen
  \bibfield  {author} {\bibinfo {author} {\bibfnamefont {F.}~\bibnamefont
  {Herrera}}\ and\ \bibinfo {author} {\bibfnamefont {F.~C.}\ \bibnamefont
  {Spano}},\ }\href {https://doi.org/10.1103/PhysRevLett.116.238301} {\bibfield
   {journal} {\bibinfo  {journal} {Phys. Rev. Lett.}\ }\textbf {\bibinfo
  {volume} {116}},\ \bibinfo {pages} {238301} (\bibinfo {year}
  {2016})}\BibitemShut {NoStop}%
\bibitem [{\citenamefont {Yang}\ and\ \citenamefont {Cao}(2021)}]{Yang2021}%
  \BibitemOpen
  \bibfield  {author} {\bibinfo {author} {\bibfnamefont {P.~Y.}\ \bibnamefont
  {Yang}}\ and\ \bibinfo {author} {\bibfnamefont {J.}~\bibnamefont {Cao}},\
  }\href {https://doi.org/10.1021/acs.jpclett.1c02210} {\bibfield  {journal}
  {\bibinfo  {journal} {J. Phys. Chem. Lett.}\ }\textbf {\bibinfo {volume}
  {12}},\ \bibinfo {pages} {9531} (\bibinfo {year} {2021})}\BibitemShut
  {NoStop}%
\bibitem [{\citenamefont {Li}\ \emph {et~al.}(2021{\natexlab{b}})\citenamefont
  {Li}, \citenamefont {Mandal},\ and\ \citenamefont {Huo}}]{Li2021a}%
  \BibitemOpen
  \bibfield  {author} {\bibinfo {author} {\bibfnamefont {X.}~\bibnamefont
  {Li}}, \bibinfo {author} {\bibfnamefont {A.}~\bibnamefont {Mandal}},\ and\
  \bibinfo {author} {\bibfnamefont {P.}~\bibnamefont {Huo}},\ }\href
  {https://doi.org/10.1038/s41467-021-21610-9} {\bibfield  {journal} {\bibinfo
  {journal} {Nat. Commun.}\ }\textbf {\bibinfo {volume} {12}},\ \bibinfo
  {pages} {1315} (\bibinfo {year} {2021}{\natexlab{b}})}\BibitemShut {NoStop}%
\bibitem [{\citenamefont {Simpkins}\ \emph {et~al.}(2021)\citenamefont
  {Simpkins}, \citenamefont {Dunkelberger},\ and\ \citenamefont
  {Owrutsky}}]{Simpkins2021}%
  \BibitemOpen
  \bibfield  {author} {\bibinfo {author} {\bibfnamefont {B.~S.}\ \bibnamefont
  {Simpkins}}, \bibinfo {author} {\bibfnamefont {A.~D.}\ \bibnamefont
  {Dunkelberger}},\ and\ \bibinfo {author} {\bibfnamefont {J.~C.}\ \bibnamefont
  {Owrutsky}},\ }\href {https://doi.org/10.1021/acs.jpcc.1c05362} {\bibfield
  {journal} {\bibinfo  {journal} {J. Phys. Chem. C}\ }\textbf {\bibinfo
  {volume} {125}},\ \bibinfo {pages} {19081} (\bibinfo {year}
  {2021})}\BibitemShut {NoStop}%
\bibitem [{\citenamefont {Philbin}\ \emph {et~al.}(2022)\citenamefont
  {Philbin}, \citenamefont {Wang}, \citenamefont {Narang},\ and\ \citenamefont
  {Dou}}]{Philbin2022}%
  \BibitemOpen
  \bibfield  {author} {\bibinfo {author} {\bibfnamefont {J.~P.}\ \bibnamefont
  {Philbin}}, \bibinfo {author} {\bibfnamefont {Y.}~\bibnamefont {Wang}},
  \bibinfo {author} {\bibfnamefont {P.}~\bibnamefont {Narang}},\ and\ \bibinfo
  {author} {\bibfnamefont {W.}~\bibnamefont {Dou}},\ }\href
  {https://doi.org/10.1021/acs.jpcc.2c04741} {\bibfield  {journal} {\bibinfo
  {journal} {J. Phys. Chem. C}\ }\textbf {\bibinfo {volume} {126}},\ \bibinfo
  {pages} {14908} (\bibinfo {year} {2022})}\BibitemShut {NoStop}%
\bibitem [{\citenamefont {White}\ \emph {et~al.}(2020)\citenamefont {White},
  \citenamefont {Gao}, \citenamefont {Minnich},\ and\ \citenamefont
  {Chan}}]{White2020}%
  \BibitemOpen
  \bibfield  {author} {\bibinfo {author} {\bibfnamefont {A.~F.}\ \bibnamefont
  {White}}, \bibinfo {author} {\bibfnamefont {Y.}~\bibnamefont {Gao}}, \bibinfo
  {author} {\bibfnamefont {A.~J.}\ \bibnamefont {Minnich}},\ and\ \bibinfo
  {author} {\bibfnamefont {G.~K.~L.}\ \bibnamefont {Chan}},\ }\href
  {https://doi.org/10.1063/5.0033132} {\bibfield  {journal} {\bibinfo
  {journal} {J. Chem. Phys.}\ }\textbf {\bibinfo {volume} {153}},\ \bibinfo
  {pages} {224112} (\bibinfo {year} {2020})}\BibitemShut {NoStop}%
\bibitem [{\citenamefont {Eisenschitz}\ and\ \citenamefont
  {London}(1930)}]{Eisenschitz1930}%
  \BibitemOpen
  \bibfield  {author} {\bibinfo {author} {\bibfnamefont {R.}~\bibnamefont
  {Eisenschitz}}\ and\ \bibinfo {author} {\bibfnamefont {F.}~\bibnamefont
  {London}},\ }\href {https://doi.org/10.1007/BF01341258} {\bibfield  {journal}
  {\bibinfo  {journal} {Zeitschrift f{\"{u}}r Phys.}\ }\textbf {\bibinfo
  {volume} {60}},\ \bibinfo {pages} {491} (\bibinfo {year} {1930})}\BibitemShut
  {NoStop}%
\bibitem [{\citenamefont {London}(1930)}]{London1930}%
  \BibitemOpen
  \bibfield  {author} {\bibinfo {author} {\bibfnamefont {F.}~\bibnamefont
  {London}},\ }\href {https://doi.org/10.1007/BF01421741} {\bibfield  {journal}
  {\bibinfo  {journal} {Zeitschrift f{\"{u}}r Phys.}\ }\textbf {\bibinfo
  {volume} {63}},\ \bibinfo {pages} {245} (\bibinfo {year} {1930})}\BibitemShut
  {NoStop}%
\bibitem [{\citenamefont {Dahlke}\ and\ \citenamefont
  {Truhlar}(2007)}]{Dahlke2007}%
  \BibitemOpen
  \bibfield  {author} {\bibinfo {author} {\bibfnamefont {E.~E.}\ \bibnamefont
  {Dahlke}}\ and\ \bibinfo {author} {\bibfnamefont {D.~G.}\ \bibnamefont
  {Truhlar}},\ }\href {https://doi.org/10.1021/ct600253j} {\bibfield  {journal}
  {\bibinfo  {journal} {J. Chem. Theory Comput.}\ }\textbf {\bibinfo {volume}
  {3}},\ \bibinfo {pages} {46} (\bibinfo {year} {2007})}\BibitemShut {NoStop}%
\bibitem [{\citenamefont {Barron}(2017)}]{CELU_paper}%
  \BibitemOpen
  \bibfield  {author} {\bibinfo {author} {\bibfnamefont {J.~T.}\ \bibnamefont
  {Barron}},\ }\href {https://doi.org/10.48550/ARXIV.1704.07483} {\bibinfo
  {title} {Continuously differentiable exponential linear units}} (\bibinfo
  {year} {2017}),\ \Eprint {https://arxiv.org/abs/1704.07483}
  {arXiv:1704.07483} \BibitemShut {NoStop}%
\bibitem [{\citenamefont {Kingma}\ and\ \citenamefont
  {Ba}(2014)}]{Adam_optimizer}%
  \BibitemOpen
  \bibfield  {author} {\bibinfo {author} {\bibfnamefont {D.~P.}\ \bibnamefont
  {Kingma}}\ and\ \bibinfo {author} {\bibfnamefont {J.}~\bibnamefont {Ba}},\
  }\href {https://doi.org/10.48550/ARXIV.1412.6980} {\bibinfo {title} {Adam: A
  method for stochastic optimization}} (\bibinfo {year} {2014}),\ \Eprint
  {https://arxiv.org/abs/1412.6980} {arXiv:1412.6980} \BibitemShut {NoStop}%
\bibitem [{\citenamefont {Paszke}\ \emph {et~al.}(2019)\citenamefont {Paszke},
  \citenamefont {Gross}, \citenamefont {Massa}, \citenamefont {Lerer},
  \citenamefont {Bradbury}, \citenamefont {Chanan}, \citenamefont {Killeen},
  \citenamefont {Lin}, \citenamefont {Gimelshein}, \citenamefont {Antiga},
  \citenamefont {Desmaison}, \citenamefont {Kopf}, \citenamefont {Yang},
  \citenamefont {DeVito}, \citenamefont {Raison}, \citenamefont {Tejani},
  \citenamefont {Chilamkurthy}, \citenamefont {Steiner}, \citenamefont {Fang},
  \citenamefont {Bai},\ and\ \citenamefont {Chintala}}]{NEURIPS2019_9015}%
  \BibitemOpen
  \bibfield  {author} {\bibinfo {author} {\bibfnamefont {A.}~\bibnamefont
  {Paszke}}, \bibinfo {author} {\bibfnamefont {S.}~\bibnamefont {Gross}},
  \bibinfo {author} {\bibfnamefont {F.}~\bibnamefont {Massa}}, \bibinfo
  {author} {\bibfnamefont {A.}~\bibnamefont {Lerer}}, \bibinfo {author}
  {\bibfnamefont {J.}~\bibnamefont {Bradbury}}, \bibinfo {author}
  {\bibfnamefont {G.}~\bibnamefont {Chanan}}, \bibinfo {author} {\bibfnamefont
  {T.}~\bibnamefont {Killeen}}, \bibinfo {author} {\bibfnamefont
  {Z.}~\bibnamefont {Lin}}, \bibinfo {author} {\bibfnamefont {N.}~\bibnamefont
  {Gimelshein}}, \bibinfo {author} {\bibfnamefont {L.}~\bibnamefont {Antiga}},
  \bibinfo {author} {\bibfnamefont {A.}~\bibnamefont {Desmaison}}, \bibinfo
  {author} {\bibfnamefont {A.}~\bibnamefont {Kopf}}, \bibinfo {author}
  {\bibfnamefont {E.}~\bibnamefont {Yang}}, \bibinfo {author} {\bibfnamefont
  {Z.}~\bibnamefont {DeVito}}, \bibinfo {author} {\bibfnamefont
  {M.}~\bibnamefont {Raison}}, \bibinfo {author} {\bibfnamefont
  {A.}~\bibnamefont {Tejani}}, \bibinfo {author} {\bibfnamefont
  {S.}~\bibnamefont {Chilamkurthy}}, \bibinfo {author} {\bibfnamefont
  {B.}~\bibnamefont {Steiner}}, \bibinfo {author} {\bibfnamefont
  {L.}~\bibnamefont {Fang}}, \bibinfo {author} {\bibfnamefont {J.}~\bibnamefont
  {Bai}},\ and\ \bibinfo {author} {\bibfnamefont {S.}~\bibnamefont
  {Chintala}},\ }in\ \href
  {http://papers.neurips.cc/paper/9015-pytorch-an-imperative-style-high-performance-deep-learning-library.pdf}
  {\emph {\bibinfo {booktitle} {Advances in Neural Information Processing
  Systems 32}}},\ \bibinfo {editor} {edited by\ \bibinfo {editor}
  {\bibfnamefont {H.}~\bibnamefont {Wallach}}, \bibinfo {editor} {\bibfnamefont
  {H.}~\bibnamefont {Larochelle}}, \bibinfo {editor} {\bibfnamefont
  {A.}~\bibnamefont {Beygelzimer}}, \bibinfo {editor} {\bibfnamefont
  {F.}~\bibnamefont {d\textquotesingle Alch\'{e}-Buc}}, \bibinfo {editor}
  {\bibfnamefont {E.}~\bibnamefont {Fox}},\ and\ \bibinfo {editor}
  {\bibfnamefont {R.}~\bibnamefont {Garnett}}}\ (\bibinfo  {publisher} {Curran
  Associates, Inc.},\ \bibinfo {year} {2019})\ pp.\ \bibinfo {pages}
  {8024--8035}\BibitemShut {NoStop}%
\bibitem [{\citenamefont {Bussi}\ and\ \citenamefont
  {Parrinello}(2007)}]{Bussi_Langevin}%
  \BibitemOpen
  \bibfield  {author} {\bibinfo {author} {\bibfnamefont {G.}~\bibnamefont
  {Bussi}}\ and\ \bibinfo {author} {\bibfnamefont {M.}~\bibnamefont
  {Parrinello}},\ }\href {https://doi.org/10.1103/PhysRevE.75.056707}
  {\bibfield  {journal} {\bibinfo  {journal} {Phys. Rev. E}\ }\textbf {\bibinfo
  {volume} {75}},\ \bibinfo {pages} {056707} (\bibinfo {year}
  {2007})}\BibitemShut {NoStop}%
\bibitem [{\citenamefont {Ceriotti}\ \emph {et~al.}(2009)\citenamefont
  {Ceriotti}, \citenamefont {Bussi},\ and\ \citenamefont
  {Parrinello}}]{Ceriotti_PRL:2009}%
  \BibitemOpen
  \bibfield  {author} {\bibinfo {author} {\bibfnamefont {M.}~\bibnamefont
  {Ceriotti}}, \bibinfo {author} {\bibfnamefont {G.}~\bibnamefont {Bussi}},\
  and\ \bibinfo {author} {\bibfnamefont {M.}~\bibnamefont {Parrinello}},\
  }\href {https://doi.org/10.1103/PhysRevLett.103.030603} {\bibfield  {journal}
  {\bibinfo  {journal} {Phys. Rev. Lett.}\ }\textbf {\bibinfo {volume} {103}},\
  \bibinfo {pages} {030603} (\bibinfo {year} {2009})}\BibitemShut {NoStop}%
\bibitem [{\citenamefont {Ceriotti}\ \emph
  {et~al.}(2010{\natexlab{a}})\citenamefont {Ceriotti}, \citenamefont {Bussi},\
  and\ \citenamefont {Parrinello}}]{Ceriotti_GLE_2010}%
  \BibitemOpen
  \bibfield  {author} {\bibinfo {author} {\bibfnamefont {M.}~\bibnamefont
  {Ceriotti}}, \bibinfo {author} {\bibfnamefont {G.}~\bibnamefont {Bussi}},\
  and\ \bibinfo {author} {\bibfnamefont {M.}~\bibnamefont {Parrinello}},\
  }\href {https://doi.org/10.1021/ct900563s} {\bibfield  {journal} {\bibinfo
  {journal} {J. Chem. Theory Comput.}\ }\textbf {\bibinfo {volume} {6}},\
  \bibinfo {pages} {1170} (\bibinfo {year} {2010}{\natexlab{a}})}\BibitemShut
  {NoStop}%
\bibitem [{\citenamefont {Ceriotti}\ \emph {et~al.}(2011)\citenamefont
  {Ceriotti}, \citenamefont {Manolopoulos},\ and\ \citenamefont
  {Parrinello}}]{Ceriotti_PI-GLE_2011}%
  \BibitemOpen
  \bibfield  {author} {\bibinfo {author} {\bibfnamefont {M.}~\bibnamefont
  {Ceriotti}}, \bibinfo {author} {\bibfnamefont {D.~E.}\ \bibnamefont
  {Manolopoulos}},\ and\ \bibinfo {author} {\bibfnamefont {M.}~\bibnamefont
  {Parrinello}},\ }\href {https://doi.org/10.1063/1.3556661} {\bibfield
  {journal} {\bibinfo  {journal} {J. Chem. Phys.}\ }\textbf {\bibinfo {volume}
  {134}},\ \bibinfo {pages} {084104} (\bibinfo {year} {2011})}\BibitemShut
  {NoStop}%
\bibitem [{\citenamefont {Ceriotti}\ \emph
  {et~al.}(2010{\natexlab{b}})\citenamefont {Ceriotti}, \citenamefont
  {Parrinello}, \citenamefont {Markland},\ and\ \citenamefont
  {Manolopoulos}}]{Ceriotti_JCP_2010}%
  \BibitemOpen
  \bibfield  {author} {\bibinfo {author} {\bibfnamefont {M.}~\bibnamefont
  {Ceriotti}}, \bibinfo {author} {\bibfnamefont {M.}~\bibnamefont
  {Parrinello}}, \bibinfo {author} {\bibfnamefont {T.~E.}\ \bibnamefont
  {Markland}},\ and\ \bibinfo {author} {\bibfnamefont {D.~E.}\ \bibnamefont
  {Manolopoulos}},\ }\href {https://doi.org/10.1063/1.3489925} {\bibfield
  {journal} {\bibinfo  {journal} {J. Chem. Phys.}\ }\textbf {\bibinfo {volume}
  {133}},\ \bibinfo {pages} {124104} (\bibinfo {year}
  {2010}{\natexlab{b}})}\BibitemShut {NoStop}%
\bibitem [{\citenamefont {Ceriotti}\ \emph {et~al.}(2014)\citenamefont
  {Ceriotti}, \citenamefont {More},\ and\ \citenamefont
  {Manolopoulos}}]{CERIOTTI20141019}%
  \BibitemOpen
  \bibfield  {author} {\bibinfo {author} {\bibfnamefont {M.}~\bibnamefont
  {Ceriotti}}, \bibinfo {author} {\bibfnamefont {J.}~\bibnamefont {More}},\
  and\ \bibinfo {author} {\bibfnamefont {D.~E.}\ \bibnamefont {Manolopoulos}},\
  }\href {https://doi.org/https://doi.org/10.1016/j.cpc.2013.10.027} {\bibfield
   {journal} {\bibinfo  {journal} {Comput. Phys. Commun.}\ }\textbf {\bibinfo
  {volume} {185}},\ \bibinfo {pages} {1019} (\bibinfo {year}
  {2014})}\BibitemShut {NoStop}%
\end{thebibliography}%

\end{document}